\documentclass[twocolumn,showpacs,preprintnumbers,amsmath,amssymb]{revtex4}

\usepackage{amsmath,amsfonts,amssymb,graphics,graphicx,epsfig,color,times,bbm}
\usepackage{esint}
\usepackage{dcolumn}

\newcommand{\op}{\widehat}

\begin{document}

\preprint{APS/123-QED}

\title{Mode-selective quantization and multimodal effective models\\
for spherically layered systems}

\author{D. Dzsotjan,$^{1,2}$ B. Rousseaux,$^1$ H. R. Jauslin,$^1$ G. Colas des Francs,$^1$ C. Couteau,$^{3,4,5}$ S. Gu\'erin$^{1.}$}
\affiliation{$^1$Laboratoire Interdisciplinaire Carnot de Bourgogne, CNRS UMR 6303, Universit\'e Bourgogne Franche-Comt\'{e},
BP 47870, 21078 Dijon, France}
\email{david.dzsotjan@u-bourgogne.fr}
\affiliation{$^2$Wigner Research Centre for Physics, Hungarian Academy of Sciences, Konkoly-Thege M. \'{u}t 29-33, H-1121 Budapest, Hungary}
\affiliation{$^3$Laboratory for Nanotechnology, Instrumentation and Optics, ICB CNRS UMR 6281, University of Technology of Troyes, 10000 Troyes, France}
\affiliation{$^4$CINTRA CNRS-NTU-Thales, UMI 3288, Singapore 637553}
\affiliation{$^5$Centre for Disruptive Photonic Technologies, Nanyang Technological University, Singapore 637371}



\date{\today}

\begin{abstract}

We propose a geometry-specific, mode-selective quantization scheme in coupled field-emitter systems which makes it easy to include material and geometrical properties, intrinsic losses as well as the positions of an arbitrary number of quantum emitters. The method is presented through the example of a spherically symmetric, non-magnetic, arbitrarily layered system.  We follow it up by a framework to project the system on simpler, effective  cavity QED models. Maintaining a well-defined connection to the original quantization, we derive the emerging effective quantities from  the full, mode-selective model in a mathematically consistent way. We discuss the uses and limitations of these effective models.

\end{abstract}

\pacs{Valid PACS appear here}
\maketitle


\section{Introduction}

Quantum technology applications heavily involve implementations of all-optical devices and information processing at the quantum level, for which it is necessary to control the interaction of quantum emitters with light. Optical resonators make up only a portion of setups where strong light-emitter interaction is achieved: quantum emitters coupled to plasmonic eigenmodes of metallic nanostructures constitute a promising class of such systems \cite{Chang2007a,Chang2006, Enoch2012, Tame2013}. Here, the emitters interact with extremely confined travelling or local plasmon modes, so that a strong coupling interaction regime is possible \cite{ Dzsotjan2010, Gonzalez-Tudela2010, Hakami2014, Delga2014a, Delga2014b, Nerkararyan2014, Waks2010, Zengin2015, Bellessa2004}.

Regardless of the actual physical realisation, all these systems essentially consist of quantum emitters interacting with bosonic resonances, thus, the multimode effective models developed in cavity quantum electrodynamics (cQED) are applicable to the whole range of them. They have the advantage to be very clear and efficient, however, assigning the values of the coupling constants is not always an easy task. 

On the other hand, there exist quantization methods based on the linear response of the environment surrounding the quantum emitters \cite{Huttner1992, Gruner1996, Dung1998, Knoll2001, Suttorp2004a, Suttorp2004b, Vogel2006, Philbin2010}. Based on the Green tensor of the system, they allow for introducing geometrical and material properties, as well as emitter positioning into the model. Also, the detailed contributions to the behaviour of a given coupled system are often difficult to resolve. 

The main goal of this paper is to create a framework that connects a mode-selective, Green tensor based quantization with the effective quantum mechanical models used in cavity QED. In this way, one can precisely calculate the emitter-field coupling constants from the geometry and arrangement of the systems, the material properties of the dielectric bodies in it, and the position of the quantum emitters (atoms, molecules, quantum dots, etc). Being able to introduce and position multiple emitters is an important feature that allows for studying their collective interactions, mediated by the modes of the environment. We give a consistent guideline for creating a geometry-specific, mode-selective quantization relying on the Green tensor of the  system. Being inherently geometry-specific, the quantization is demonstrated through the example of a spherically layered, non-magnetic medium, but it can be analogously extended to geometries with different symmetries. 

From this mode-selective, quantized model we derive a class of effective cQED models where the field operators depend only on frequency and the mode indices. As a result, one can work with a much more streamlined, easier-to-handle model that only contains the parameters which are necessary for the given investigation. Applying a further transformation, one even has the freedom to keep track only of the relevant mode indices and get rid of the others. The transition to these effective models  is performed by the aid of a separation into "dark" and "bright" subspaces of the total Hilbert space.
 
 Finally, we present a way to derive a class of effective cQED models where the creation/annihilation operators no longer have a continuous spectral dependence - instead, they each are globally associated to respective resonance peaks of the structured reservoir. It is a useful model if one is not interested in following the excitations in the field in a frequency-resolved fashion, the only relevant information being which resonance peak is excited as a whole. Here, each separate mode is represented by a single, lossy field operator.

The connection with the original, mode-selective quantization is well defined, so that all the required parameters in the effective models can be easily obtained from the Green tensor of the system.  

The article is organized as follows. In Sec. \ref{sec:mode_exp_sph} we set the stage for the quantization by expanding the electromagnetic field in a spherically symmetric, nonmagnetic, multilayer system and describing the structure of the Green tensor. Sec. \ref{sec:mode-selective_quant} presents the mode-selective quantization scheme where we derive operators associated with particular vector harmonic terms of the field. Next, in Sec. \ref{sec:effective_models} different kinds of effective cQED models are presented, as well as their connection to the mode-selective quantization. We derive a continuous effective model with a single (Sec. \ref{sec:effective_cont_single}), as well as several emitters (Sec. \ref{sec:effective_cont_multiple}). This is followed by the derivation of discrete effective models where we have a single field operator per resonance, for a single, as well as for several emitters in Secs. \ref{sec:effective_discrete_single} and \ref{sec:effective_discrete_multiple}, respectively. We 
summarize 
the results and implications in Sec. \ref{sec:summary}.

\section{Classical mode expansion in a spherically symmetric, multilayer system}\label{sec:mode_exp_sph}

In the following, we are going to show that it is possible to quantize a spherically symmetric system in a way where the field operators create and annihilate elementary excitations associated with individual spherical harmonic orders. The initial steps of the method follow the scheme presented in \cite{Dung1998}. We start with Maxwell's classical equations, where we introduce a charge and current density source term $\rho_N$ and $\vec{j}_N$, respectively, usually called ``noise polarisation`` and ``noise current'' in the literature. They are needed in order to later construct creation and annihilation operators for elementary excitations. In Fourier space, we can write
  \begin{eqnarray}
 \boldsymbol{\nabla}\cdot\mathbf{B}(\mathbf{r},\omega)&=&0\\
 \boldsymbol{\nabla}\times\mathbf{E}(\mathbf{r},\omega)&=&\mathrm{i}\omega\mathbf{B}(\mathbf{r},\omega)\label{eq:Maxwell2}\\ 
 \epsilon_0\boldsymbol{\nabla}\cdot\epsilon(\mathbf{r},\omega)\mathbf{E}(\mathbf{r},\omega)&=&\rho_N(\mathbf{r},\omega)\\
 \boldsymbol{\nabla}\times\mathbf{B}(\mathbf{r},\omega)+\mathrm{i}\frac{\omega}{c^2}\epsilon(\mathbf{r},\omega)\mathbf{E}(\mathbf{r},\omega)&=&\mu_0\mathbf{j}_N(\mathbf{r},\omega)\label{eq:Maxwell4}
\end{eqnarray}
where the noise charge density and the noise current density, respectively, are
\begin{eqnarray}
 \rho_N(\mathbf{r},\omega)=-\boldsymbol{\nabla}\cdot\mathbf{P}_N(\mathbf{r},\omega)\\
 \mathbf{j}_N(\mathbf{r},\omega)=-\mathrm{i}\omega\mathbf{P}_N(\mathbf{r},\omega).
\end{eqnarray}
The noise polarisation $\mathbf{P}_N$ shows up as a small fluctuation term in the polarisation:
\begin{equation}
 \mathbf{P}(\mathbf{r},\omega)=\epsilon_0[\epsilon(\mathbf{r},\omega)-1]\mathbf{E}(\mathbf{r},\omega)+\mathbf{P}_N(\mathbf{r},\omega).
\end{equation}

Combining Eqs (\ref{eq:Maxwell2}) and (\ref{eq:Maxwell4}), we get the wave equation for the electric field, with the noise current as source:
\begin{equation}
 \boldsymbol{\nabla}\times\boldsymbol{\nabla}\times\mathbf{E}(\mathbf{r},\omega)-\frac{\omega^2}{c^2}\epsilon(\mathbf{r},\omega)\mathbf{E}(\mathbf{r},\omega)=\mathrm{i}\omega\mu_0\mathbf{j}_N(\mathbf{r},\omega).
\end{equation}
In terms of the Green tensor of the system, the classical electric field reads
\begin{equation}
\mathbf{E}(\mathbf{r},\omega)=\mathrm{i}\omega\mu_0\int\!\!\mathrm{d}^3r^\prime\bar{\bar{G}}(\mathbf{r},\mathbf{r^\prime},\omega)\cdot\mathbf{j}_N(\mathbf{r^\prime},\omega)\label{eq:E}
\end{equation}
where the Green tensor obeys the Maxwell-Helmholtz wave equation:
\begin{equation}
 \boldsymbol{\nabla}\times\boldsymbol{\nabla}\times\bar{\bar{G}}(\mathbf{r},\mathbf{r^\prime},\omega)-\frac{\omega^2}{c^2}\epsilon(\mathbf{r},\omega)\bar{\bar{G}}(\mathbf{r},\mathbf{r^\prime},\omega)=\bar{\bar{\delta}}(\mathbf{r}-\mathbf{r^\prime}).\label{eq:M-H}
\end{equation}

So far, the description is completely independent of the specific geometry of any given system. However, in order to construct a class of operators which separately create or annihilate excitations in arbitrary modes of a system, we have to take into account the geometry inherent to it. 

The Green tensor for a given system can be expanded on the tensorial basis obtained from a well chosen set of harmonic vector functions.  On the one hand, this serves the technical purpose of fulfilling the boundary conditions. On the other hand, if the system has intrinsic resonances, using this expansion enables one to efficiently describe and handle these, the harmonic vector functions containing the properties of the current geometry. In this work, we consider a spherically layered geometry, and, as a special example, a spherical metallic nanoparticle, which is a benchmark configuration for a cQED-like description of localized surface plasmons.
Solving the equation
\begin{equation}
 \left(\nabla^2+q^2\right)\psi_{nm^e_o }(\mathbf{r},q)=0,
\end{equation}
we obtain, as described in \cite{Li1994, Tai1993, Chew1995}, the spherical vector harmonic eigenfunctions:
\begin{equation}
 \psi_{ nm^e_o}(\mathbf{r},q)=z_n(qr)P^m_n(\cos\theta)\begin{array}{c}
                                                               \cos \\ \sin
                                                              \end{array}(m\phi),\label{eq:scalar_harmonics}
\end{equation}
where $q\ge 0$ is a parameter having the dimensionality of the wave number, $e$ and $o$ refer to ``even'' and ``odd'', respectively, and $n$, $m$ are discrete harmonic indices. $z_n$ denotes a spherical harmonic function of $n$-order which can be a spherical Bessel or a spherical Hankel function, depending on regularisation requirements. Finally, $P^m_n$ are the associated Legendre polynomials, labeled with the harmonic indices. The spherical vector harmonics used for the expansion of the Green tensor are constructed as
\begin{eqnarray}
 \begin{aligned}
\mathbf{M}_{nm^e_o }(\mathbf{r},q)&=\boldsymbol{\nabla}\times\left[\psi_{nm^e_o }(\mathbf{r},q)\mathbf{r}\right]\\
\mathbf{N}_{nm^e_o }(\mathbf{r},q)&=\frac{1}{q}\boldsymbol{\nabla}\times\boldsymbol{\nabla}\times\left[\psi_{nm^e_o }(\mathbf{r},q)\mathbf{r}\right]\\
\mathbf{L}_{nm^e_o }(\mathbf{r},q)&=\boldsymbol{\nabla}\psi_{nm^e_o }(\mathbf{r},q) \label{eq:vector_harmonics},
 \end{aligned}
\end{eqnarray}

where $\mathbf{M}(\mathbf{r},q)$ and $\mathbf{N}(\mathbf{r},q)$ are continuous-spectrum eigenvectors of
\begin{equation}
 \boldsymbol{\nabla}\times\boldsymbol{\nabla}\times\mathbf{K}(\mathbf{r},q)=q^2\mathbf{K}(\mathbf{r},q)
\end{equation}
with eigenvalue $q^2$, and $\mathbf{L}(\mathbf{r},q)$ spans the nullspace of the $\boldsymbol{\nabla}\times\boldsymbol{\nabla}\times$ operator. It can be shown that the spherical vector harmonics form a complete basis parametrized by the discrete indices $n=0,1,2,...$ (to $\infty$), $m=0,1,2,...,n$ , $p=e,o$ (\emph{even} or \emph{odd}), as well as the continuous variable $q \ge 0$. The basis is orthogonal in these parameters but it is not normalized. Appendix \ref{sec:app_spherical_harm} contains the explicit form of the spherical vector harmonics in detail, as well as their orthogonality relations.

Due to the layered nature of the system, we will also encounter the situation of $q$ depending on $r$ (i.e., the radius). In this case, the orthogonality relations change to
\begin{eqnarray}
 \begin{aligned}
\int\!\!\mathrm{d^3}r \mathbf{K}_{nmp }(\mathbf{r},q(\mathbf{r}))\cdot\mathbf{K}_{n^\prime m^\prime p^\prime}(\mathbf{r},q^\prime(\mathbf{r}))\quad\quad\quad&\\
=S^K_{n}(q,q^\prime)\,\,\mathcal{Q}^K_{nmp}\delta_{nn^\prime}\delta_{mm^\prime}&\delta_{pp^\prime},\\
\label{eq:ortho3}
 \end{aligned}
\end{eqnarray}
where $\mathbf{K}=\mathbf{M},\mathbf{N},\mathbf{L}$ and
\begin{equation}
 S^K_{n}(q,q^\prime)=\int_0^\infty\!\!\!\!\!\mathrm{d}r\, r^2\! K_{n}[q(r)r]K_{n}[q^\prime(r)r],
\end{equation}
 the integrand 	being the $r$ - dependent expression that is left after one performs the integrations over the angles $\theta$ and $\phi$. However, the changes due to the layered structure do not affect the orthogonality between the spherical indices $n,n^\prime$, $m,m^\prime$ and the parity $p=e,o$. The normalization factor reads
\begin{equation}
 \mathcal{Q}^K_{nm^e_o}=\left\{ \begin{array}{l l} \frac{2\pi n(n+1)(n+m)!}{ (2n+1)(n-m)!}(1 \pm \delta_{m0}), & K=M,N \\
					      \frac{2\pi (n+m)!}{ (2n+1)(n-m)!}(1 \pm \delta_{m0}), & K=L
                         \end{array}  \right.   ,\label{eq:orth_norm_factor}
\end{equation}
the upper and lower signs on the right hand side referring to even and odd parities, respectively.

\begin{figure}
 \includegraphics[width=0.45\textwidth]{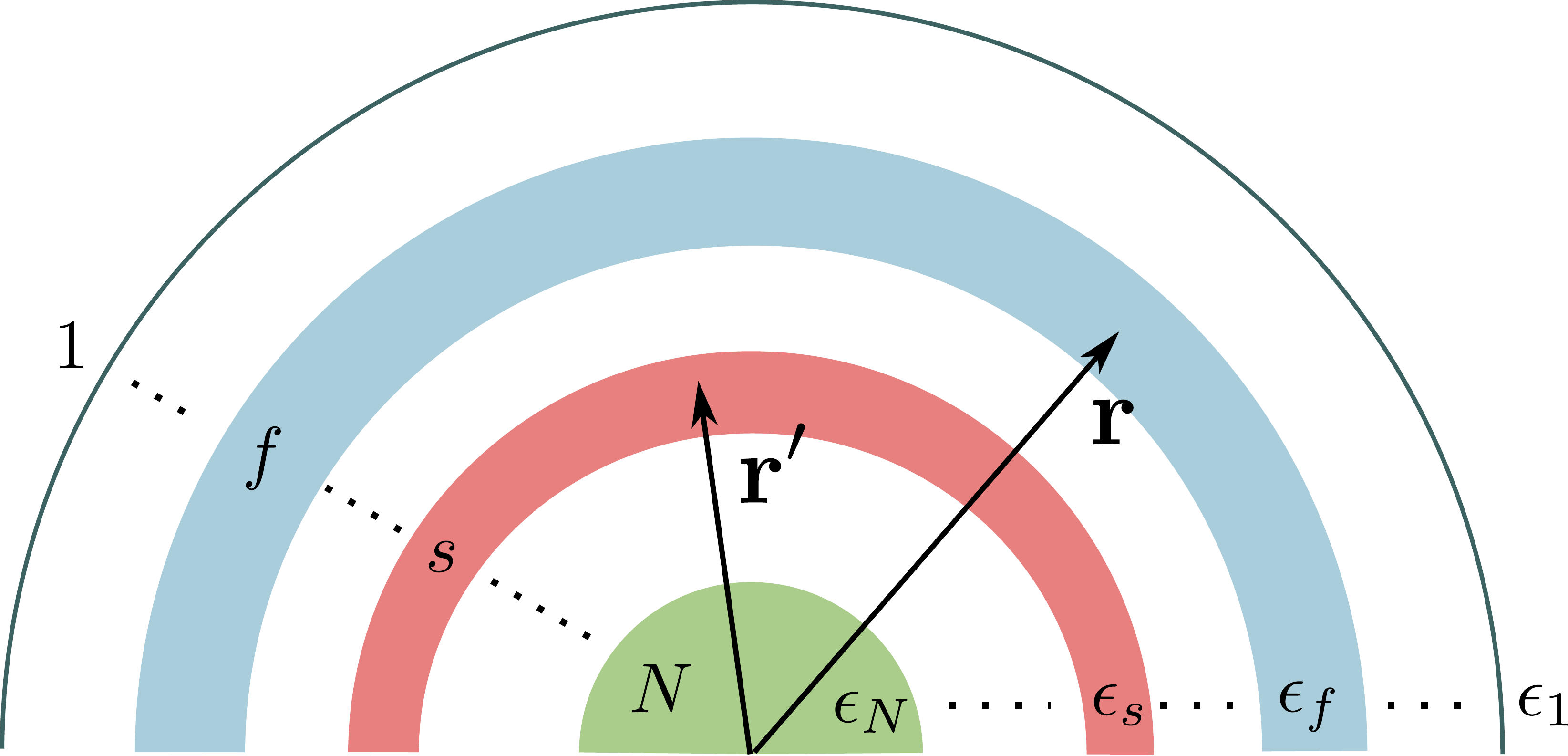}
 \caption{Structure of a spherically N-layered medium, where the material properties are piecewise homogeneous between individual layer interfaces. The layers containing the field ($\mathbf{r}$) and the source ($\mathbf{r^\prime}$) points are labelled with $f$ and $s$, respectively.  \label{fig:layers}}
\end{figure}

The Green tensor of a general $N$-layered spherical system (shown in Fig. \ref{fig:layers} ) is expressed as
\begin{equation}
 \bar{\bar{G}}(\mathbf{r},\mathbf{r^\prime},\omega)=\bar{\bar{G}}_0(\mathbf{r},\mathbf{r^\prime},\omega)\delta_{fs}+\bar{\bar{G}}_S(\mathbf{r},\mathbf{r^\prime},\omega),
\end{equation}
where $f$ is the index of the layer wherein the field point $\mathbf{r}$ is located, and the source point $\mathbf{r^\prime}$ is contained in layer $s$. In case $\mathbf{r}$ and $\mathbf{r^\prime}$ are in the same layer, we have a term that can be interpreted as direct propagation between source and field point ($\bar{\bar{G}}_0$) and another that results from scattering on the surrounding layers ($\bar{\bar{G}}_S$). If $f\neq s$, only the scattered term is present.

For a spherically layered system, in order to fulfill the boundary conditions on the layer interfaces, we expand Green's tensor in the tensorial basis of spherical vector harmonics (see Appendix \ref{sec:app_Greens_tensor} for details),
\begin{equation}
 \bar{\bar{G}}(\mathbf{r},\mathbf{r^\prime},\omega)=\sum_{n=0}^{\infty}\sum_{m=0}^{n}\sum_{p=e,o}\bar{\bar{G}}^{(nmp)}(\mathbf{r},\mathbf{r^\prime},\omega),\label{eq:G}
\end{equation}
the expansion terms having the general form
\begin{eqnarray}
 \begin{aligned}
 \bar{\bar{G}}^{(nmp)}&(\mathbf{r},\mathbf{r^\prime},\omega)
 \\=\,\,&\frac{1}{k_s^2}\delta_{fs}\,\,\bar{\bar{\delta}}^{(nmp)}(\mathbf{r}-\mathbf{r^\prime})\\
 \!\!+&\sum_{j,l=0,1}\!\!\!\big[D^{jl}_{nmp}(\omega)\mathbf{M}^{(j)}_{nmp}(\mathbf{r},k_f)\otimes\mathbf{M}^{(l)}_{nmp}(\mathbf{r^\prime},k_s) \\
 &\quad\,\,\,\,\, +E^{jl}_{nmp}(\omega)\mathbf{N}^{(j)}_{nmp}(\mathbf{r},k_f)\otimes\mathbf{N}^{(l)}_{nmp}(\mathbf{r^\prime},k_s)\big],\label{eq:G_expansion}
 \end{aligned}
\end{eqnarray}
where 
\begin{eqnarray}
\begin{aligned}
 \bar{\bar{\delta}}^{(nmp)}&(\mathbf{r}-\mathbf{r^\prime})\\
 &=\int_0^\infty\!\!\!\!\!\mathrm{d}q\,\, C_{nmp}(q)\big[\mathbf{N}^{(0)}_{nmp}(\mathbf{r},q)\otimes\mathbf{N}^{(0)}_{nmp}(\mathbf{r^\prime},q)\\
  &\quad\quad\,\,\,+\mathbf{L}^{(0)}_{nmp}(\mathbf{r},q)\otimes\mathbf{L}^{(0)}_{nmp}(\mathbf{r^\prime},q)n(n+1)\big]_{\mathbf{r}\otimes\mathbf{r}}
  \end{aligned}
\end{eqnarray}
is the expansion term of the singular term in $\bar{\bar{G}}_0$, so that
\begin{equation}
 \sum_{nmp}\bar{\bar{\delta}}^{(nmp)}(\mathbf{r}-\mathbf{r^\prime})=\delta(\mathbf{r}-\mathbf{r^\prime})\hat{\mathbf{r}}\otimes\hat{\mathbf{r}},
\end{equation}
as explained in Appendix \ref{sec:app_Dirac}. Note that the integral contains the tensorial products of the \emph{radial} components of $\mathbf{N}$ and $\mathbf{L}$ only. The indices $j$ and $l$ can assume the values $0$ and $1$. For the value $0$, we construct the vector harmonics with replacing the radial function $z_n(qr)$ in (\ref{eq:scalar_harmonics}) by a spherical Bessel function of the first kind ($j_n(qr)$), and for the value $1$, we replace it by a spherical Hankel function of the first kind ($h^{(1)}_n(qr)$). The expansion coefficients $D$ and $E$ are chosen so that the boundary conditions are fulfilled and, if $f=s$, they contain the coefficients of the direct contribution ($\bar{\bar{G}}_0$) as well (see Appendix \ref{sec:app_Greens_tensor}). We assume the separate layers piecewise homogeneous, thus 
\begin{equation}
 k_{f,s}=\sqrt{\epsilon_{f,s}}\frac{\omega}{c},
\end{equation}
where $\epsilon_{f,s}$ is the relative electric permittivity corresponding to the layer $f$ and $s$, respectively. 

We can create a similar expansion for the noise current as well, so that instead of having a globally defined $\mathbf{j}_N$, we can manage separate currents labelled by the mode expansion indices $n$, $m$ and the parity $p$. 
\begin{equation}
 \mathbf{j}_N(\mathbf{r},\omega)=\sum_{n=0}^\infty\sum_{m=0}^n\sum_{p=e,o}\mathbf{j}_N^{(nmp)}(\mathbf{r},\omega).\label{eq:j_N}
\end{equation}
Next, we split off appropriate factors in order to create so-called fundamental dynamical variables \cite{Dung1998, Philbin2010}:
\begin{equation}
\mathbf{j}_N^{(nmp)}(\mathbf{r},\omega)=\omega\sqrt{\frac{\hbar\epsilon_0}{\pi}\epsilon^{\prime\prime}(\mathbf{r},\omega)}\mathbf{f}^{(nmp)}(\mathbf{r},\omega),\label{eq:f_expansion1}
\end{equation}
where $\epsilon^{\prime\prime}$ is the imaginary part of the relative electric permittivity and
\begin{eqnarray}
 \begin{aligned}
 \mathbf{f}^{(nmp)}&(\mathbf{r},\omega)=\int_0^\infty\!\!\!\!\mathrm{d}q\big[a_{nmp}(\omega,q)\mathbf{M}^{(0)}_{nmp}(\mathbf{r},q)\\
 +&b_{nmp}(\omega,q)\mathbf{N}^{(0)}_{nmp}(\mathbf{r},q)+c_{nmp}(\omega,q)\mathbf{L}^{(0)}_{nmp}(\mathbf{r},q)\big],\label{eq:f_expansion2}
 \end{aligned}
\end{eqnarray}
that is, we represent each noise current term on the subspace spanned by the basis functions belonging to the respective $n$ and $m$ and $p$ parameters. The choice to use Bessel functions for $z_n(qr)$ in the expansion is justified because it yields an orthogonal, complete set of vectorial functions, regular at the origin.

Having constructed the expansions for $\bar{\bar{G}}$ and $\mathbf{j}_N$, we substitute (\ref{eq:G}) and (\ref{eq:j_N}) into (\ref{eq:E}), getting
\begin{equation}
 \mathbf{E}(\mathbf{r},\omega)=\mathrm{i}\omega\mu_0 \sum_{\bar{n}}\sum_{\bar{n}^\prime} \int\!\!\mathrm{d}^3r^\prime \,\,\bar{\bar{G}}^{(\bar{n})}(\mathbf{r},\mathbf{r^\prime},\omega)\cdot \mathbf{j}_{N}^{(\bar{n}^\prime)}(\mathbf{r}^\prime,\omega),\label{eq:E2}
\end{equation}
where, for readability, we have grouped the harmonic and parity indices as a vectorial parameter $\bar{n}=(n,m,p)$.
Let us recall the identity for a tensor product of two vectors, multiplied by a third one:
\begin{equation}
 (\mathbf{a}\otimes\mathbf{b})\cdot\mathbf{c}=\mathbf{a}(\mathbf{b}\cdot\mathbf{c}).\label{eq:3product}
\end{equation}
As a final step, we substitute (\ref{eq:G_expansion}), (\ref{eq:f_expansion1}) and (\ref{eq:f_expansion2}) into (\ref{eq:E2}), applying relation (\ref{eq:3product}) and using the orthogonality relations in Appendix (\ref{sec:app_spherical_harm}) and (\ref{eq:ortho3}). Note that because of the spherical symmetry of the system, $\epsilon(\mathbf{r},\omega)$ depends only radially on $\mathbf{r}$. Integrating over $\mathbf{r^\prime}$ in (\ref{eq:E2}), the wave number $k_s$ depends on which layer actually $\mathbf{r^\prime}$ is  in, and thus it changes as the radial coordinate steps across layer boundaries. However, relations (\ref{eq:ortho3}) ensure the orthogonality with respect to the harmonic indices $n$ and $m$ and parity $p$. All these considered, we arrive at an expression where the harmonic index cross-terms vanish:
\begin{equation}
 \mathbf{E}(\mathbf{r},\omega)=\mathrm{i}\omega\mu_0 \sum_{\bar{n}}\! \!\int\!\!\mathrm{d}^3r^\prime \,\,\bar{\bar{G}}^{(\bar{n})}(\mathbf{r},\mathbf{r^\prime},\omega)\cdot \mathbf{j}_{N}^{(\bar{n})}(\mathbf{r}^\prime,\omega),\label{eq:E3}
\end{equation}
thus, we have appointed independent noise currents to each of the harmonic terms of Green's tensor.

\section{Mode-selective quantization}\label{sec:mode-selective_quant}

Our aim is to quantize the combined field-matter system in a way that the creation and annihilation operators toggle excitations corresponding to the individual harmonic orders. In order to do so, we take the quantization scheme of \cite{Dung1998, Philbin2010} as a starting point and then define new operators in relation to it. 

Thus, taking the total noise current as in (\ref{eq:j_N}), we initially define the dynamic vector variable $\mathbf{f}(\mathbf{r},\omega)$ as
\begin{equation}
\mathbf{j}_N(\mathbf{r},\omega)=\omega\sqrt{\frac{\hbar\epsilon_0}{\pi}\epsilon^{\prime\prime}(\mathbf{r},\omega)}\mathbf{f}(\mathbf{r},\omega),
\end{equation}
and we subsequently derive operators from it which create and annihilate elementary excitations of the combined field-matter system and obey the following bosonic commutation relations:
\begin{eqnarray}
 \begin{aligned}
 \left[\op{f}_j(\mathbf{r},\omega),\op{f}^\dagger_l(\mathbf{r^\prime},\omega^\prime)\right]&=\delta_{jl}\delta(\mathbf{r}-\mathbf{r^\prime})\delta(\omega-\omega^\prime)\\
 \big[\op{f}_j(\mathbf{r},\omega),\op{f}_l(\mathbf{r^\prime},\omega^\prime)\big]&=\left[\op{f}^\dagger_j(\mathbf{r},\omega),\op{f}^\dagger_l(\mathbf{r^\prime},\omega^\prime)\right]=0,\label{eq:comm_rel_f}
 \end{aligned}
\end{eqnarray}
the individual vectorial components referring to polarization directions. Accordingly, the electric field operator is given by
\begin{equation}
 \op{\mathbf{E}}(\mathbf{r},\omega)=\mathrm{i}\sqrt{\frac{\hbar}{\pi\epsilon_0}}\int\!\!\mathrm{d^3}r^\prime\frac{\omega^2}{c^2}\sqrt{\epsilon^{\prime\prime}(\mathbf{r^\prime},\omega)}\bar{\bar{G}}(\mathbf{r},\mathbf{r^\prime},\omega)\cdot\op{\mathbf{f}}(\mathbf{r^\prime},\omega),\label{eq:E_op1}
\end{equation}
and the Hamiltonian for the field reads
\begin{equation}
 \op{H}_F=\int\!\!\mathrm{d^3}r\!\int_0^\infty\!\!\!\!\!\mathrm{d}\omega\,\hbar\omega\,\op{\mathbf{f}}^\dagger(\mathbf{r},\omega)\cdot\op{\mathbf{f}}(\mathbf{r},\omega).\label{eq:H_f}
\end{equation}

In order to address individual harmonic excitations separately, we define operators based on the expansion (\ref{eq:f_expansion2}):
\begin{eqnarray}
 \begin{aligned}
\op{\mathbf{f}}&(\mathbf{r},\omega)=\sum_{K=M,N,L}\sum_{mnp}\!\int_0^\infty\!\!\!\!\!\mathrm{d}q\,\,\,\frac{\mathbf{K}_{nmp}(\mathbf{r},q)}{\sqrt{Q^K_{nm}(q)}}\,\,\,\op{F}^K_{nmp}(q,\omega),\label{eq:f_op_exp}
 \end{aligned}
\end{eqnarray}
where $\op{F}^{M,N,L}$ are the mode-selective annihilation operators associated with the vectorial harmonics $\mathbf{M}$, $\mathbf{N}$, and $\mathbf{L}$, respectively. Note that they are scalar operators. Since $\mathbf{K}_{n0o}(\mathbf{r},q)=0$, we can eliminate the parity dependence of the normalization, thus $Q^K_{nm}(q)$ is defined as
\begin{equation}
 Q^K_{nm}(q)=\left\{\begin{array}{c l}\frac{\pi^2 n(n+1)(n+m)!}{q^2(2n+1)(n-m)!}(1+\delta_{m0}) & \quad K=M,N\\
					\frac{\pi^2 (n+m)!}{q^2(2n+1)(n-m)!}(1+\delta_{m0}) & \quad K=L \end{array}\right. .\label{eq:app_Q_nmq}
\end{equation}
One can interpret the parameters $\{q,n,m,p\}$ as coordinates in a spherical reciprocal space, analogous to $\{k_x,k_y,k_z\}$ of the reciprocal space in a Cartesian frame of reference.

Using (\ref{eq:f_op_exp}), we can establish the commutation relations for the mode-selective operators. Recalling the relations of Appendix (\ref{sec:app_spherical_harm}), we invert (\ref{eq:f_op_exp}) and find
\begin{eqnarray}
 \begin{aligned}
\op{F}^K_{nmp}(q,\omega)&=\frac{1}{\sqrt{Q^K_{nm}(q)}}\int\!\!\mathrm{d^3}r\,\op{\mathbf{f}}(\mathbf{r},\omega)\cdot\mathbf{K}_{nmp}(\mathbf{r},q).
 \end{aligned}
\end{eqnarray}
Subsequently, the commutation relations are easily obtained:
\begin{eqnarray}
\begin{aligned}
 \left[\op{F}^K_{nmp}(q,\omega),\op{F}^{K\dagger}_{n^\prime m^\prime p^\prime}(q^\prime,\omega^\prime)\right]=\delta_{nn^\prime}\delta_{m m^\prime}\delta_{p p^\prime}\\ 
 \times\delta(q-q^\prime)\delta(\omega-\omega^\prime)\label{eq:comm_rel_F1}
 \end{aligned}
\end{eqnarray}
with $K=M, N, L$. 

For any other combination, regardless of indices and arguments, we have
\begin{eqnarray}
 \begin{aligned}
  \left[ \op{F}^{M,N,L},\op{F}^{M,N,L}\right]=\left[ \op{F}^{M,N,L\dagger},\op{F}^{M,N,L\dagger}\right]&=0\\
  \left[ \op{F}^M,\op{F}^{N\dagger}\right]=\left[\op{F}^M,\op{F}^{L\dagger}\right]=\left[\op{F}^N,\op{F}^{L\dagger}\right]&=0.
 \end{aligned}\label{eq:comm_rel_F2}
\end{eqnarray}
We have thus specified creation and annihilation operators associated with the spherical harmonic orders. We express the field Hamiltonian (\ref{eq:H_f}) with the mode-selective operators, obtaining
\begin{eqnarray}
 \op{H}^{MS}_F=\!\!\!\sum_{K=M,N,L}\!\sum_{\bar{n}}\!\int_0^\infty\!\!\!\!\!\!\mathrm{d}q\!\int_0^\infty\!\!\!\!\!\!\mathrm{d}\omega\,\,\hbar\omega\,\,\op{F}^{K\dagger}_{\bar{n}}(q,\omega)\op{F}^K_{\bar{n}}(q,\omega).\nonumber\\
\end{eqnarray}

Let us now look at the situation where $N_e$ two-level quantum emitters, with quantum states labelled as $\{|g\rangle_j,|e\rangle_j\}$, where $j=1,...,N_e$, interact with the mode structure of the spherically layered medium. The system under consideration is then described by the following Hamiltonian in the rotating-wave approximation:
\begin{equation}
 \op{H}=\op{H}_F+\sum_{j=1}^{N_e}\hbar\omega_j\op{\sigma}^j_{ee}-\sum_{j=1}^{N_e}\int_0^\infty\!\!\!\!\!\mathrm{d}\omega\left(\op{\mathbf{E}}(\mathbf{r}_j,\omega)\cdot\mathbf{d}^j\op{\sigma}_+^j+\mathrm{H.c.}\right),\label{eq:H_RWA}
\end{equation}
where $\op{H}_F$ is the field Hamiltonian (\ref{eq:H_f}), $\omega_j$ is the resonance frequency of the $j$-th emitter, with $\op{\sigma}^j_{ee}=|e\rangle_j\langle e|$ being the projector on the excited state, and the energy of the ground state has been taken as $0$. In the interaction term, we take the electric field operator at the emitter position $\mathbf{r}_j$, $\op{\sigma}_+^j=|e\rangle_j\langle g|$ flips the quantum state of the $j$-th emitter from ground to its excited state, the transition having a dipole strength $\mathbf{d}^j$.

Combining (\ref{eq:H_f}), (\ref{eq:E_op1}) and (\ref{eq:f_op_exp}) with (\ref{eq:H_RWA}), we can construct a mode-selective Hamiltonian, where the field operators address excitations associated with spherical harmonic orders:
\begin{eqnarray}
 \begin{aligned}
 \op{H}_{M\!S}&=\sum_K\sum_{\bar{n}}\!\int_0^\infty\!\!\!\!\!\!\mathrm{d}q\!\int_0^\infty\!\!\!\!\!\!\mathrm{d}\omega\,\,\hbar\omega \,\,\op{F}^{K\dagger}_{\bar{n}}(q,\omega)\op{F}^K_{\bar{n}}(q,\omega)\\
 &+\sum_{j=1}^{N_e}\hbar\omega_j\op{\sigma}^j_{ee}-\mathrm{i}\hbar\sum_K\sum_{\bar{n}}\sum_{j=1}^{N_e}\!\int_0^\infty\!\!\!\!\!\!\!\mathrm{d}q\!\int_0^\infty\!\!\!\!\!\!\!\mathrm{d}\omega \\
 &\times\bigg[\mathbf{d}^j\!\cdot\!\mathbf{V}_K^{(\bar{n})}(\mathbf{r}_j,\omega,q)\op{F}^K_{\bar{n}}(q,\omega)\op{\sigma}_+^j-\mathrm{H.c.}\bigg], \label{eq:H_MS}
 \end{aligned}
\end{eqnarray}
with
\begin{eqnarray}
 \begin{aligned}
  \mathbf{V}_K^{\bar{n}}(\mathbf{r}_j,\omega,q)&=\frac{\omega^2}{c^2}\frac{1}{\sqrt{\hbar\pi\epsilon_0\,\, Q^K_{nm}(q)}}\\
&\times\!\!\int\!\!\!\mathrm{d}^3r\sqrt{\epsilon^{\prime\prime}(\mathbf{r},\omega)}\,\,\bar{\bar{G}}^{(\bar{n})}(\mathbf{r}_j,\mathbf{r},\omega)\cdot\mathbf{K}_{\bar{n}}(\mathbf{r},q).
 \end{aligned}
\end{eqnarray}

Thus, we have constructed a theoretical framework that allows for addressing harmonic orders individually. By construction, it is also consistent with previous, established methods of quantization mentioned in the Introduction.

\section{Effective cavity QED models}\label{sec:effective_models}

 In the preceding section, we have demonstrated that it is possible to quantize the spherically layered system with a set of field operators that selectively toggle excitations associated with the separate harmonic modes. This, in turn, allows for transposing cQED concepts to describe a range of systems where quantum emitters interact with eigenmodes, with a special regard to the field of nanophotonics and plasmonics.
 
\begin{figure}[h!]
  \includegraphics[angle=0,width=0.45\textwidth]{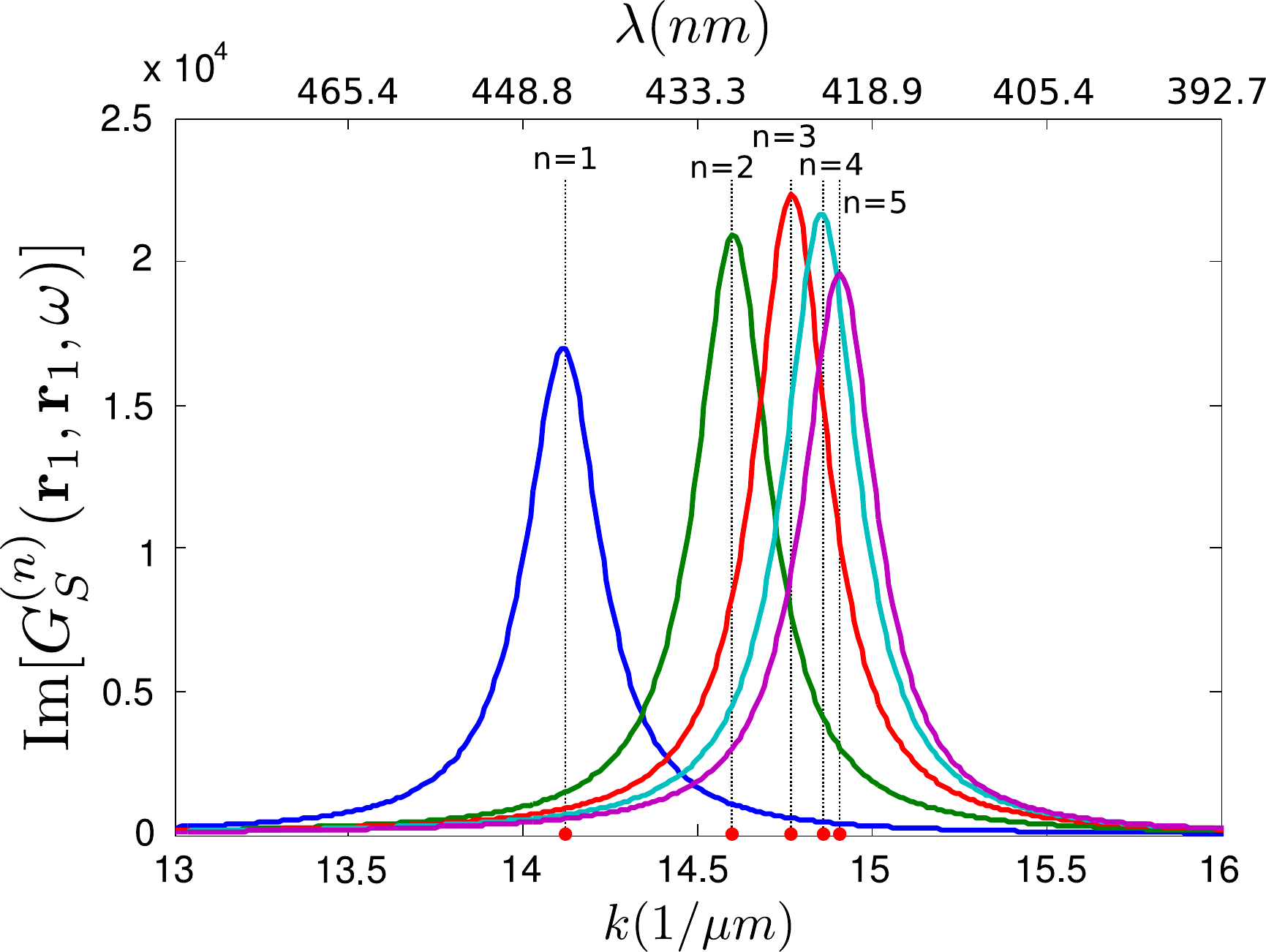}
  \caption{Scattered radial orders, i.e., local density of states (LDOS), induced by the presence of a silver nanosphere of $8$ nm radius, for an emitter with a radial dipole moment ($\mathbf{d}=d\,\hat{r}_1$), as a function of the wave number in vacuum, i.e. $k=\omega/c$. The upper axis indicates the corresponding wavelengths in units of $\mathrm{nm}$. The emitter is close to the metal surface ($r_1=10\mathrm{nm}$). Each harmonic order contains a single resonance peak.   \label{fig:scattered_orders}}
 \end{figure}

 Due to the symmetry of the geometry, resonances of such a system are structured with respect to the spherical harmonic orders. As an example, Fig. \ref{fig:scattered_orders} shows the radially projected local density of states (LDOS), i.e., $\hat{\mathbf{r}}\cdot\mathrm{Im}[\bar{\bar{G}}^{(n)}_S(\mathbf{r},\mathbf{r},\omega)]\cdot\hat{\mathbf{r}}$, due to the presence of a silver sphere, at a position close to the surface.
 
 In the calculations, similarly to \cite{Hakami2014}, we used a generalized Drude model for the relative electric permittivity of the silver nanosphere, i.e.,
 \begin{equation}
  \epsilon(\omega)=\epsilon_\infty-\frac{\omega_p^2}{\omega^2+\mathrm{i}\gamma_e\omega},
 \end{equation}
where $\epsilon_\infty$ is the high-frequency limit of the dielectric function, $\omega_p$ is the bulk plasmon frequency of the metal, and $\gamma_p$ is the Landau damping constant. Taking typical values $\epsilon_\infty=6$, $\hbar\omega_p=7.9 \mathrm{eV}$ and $\gamma_e=51\mathrm{meV}$ (see \cite{Vlack2012}) yields a good fit to experimental data (\cite{Johnson1972}) for photon energies up to $3\mathrm{eV}$.

%
 One sees immediately that each plasmonic resonance peak is associated with a separate spherical order $n$.  Thus, it is advantageous to derive cavity QED models from the full Hamiltonian (\ref{eq:H_MS}), with field operators associated to harmonic indices. In the following, we will construct cQED Hamiltonians where field operators no longer depend on the $q$ parameter and the label $K$, and only contain the relevant harmonic indices.

\subsection{Single emitter: dark and bright modes}\label{sec:effective_cont_single}

Let us consider a single quantum emitter at position $\mathbf{r_1}$ interacting with the mode structure of a spherically layered, nonmagnetic medium, i.e., $N_e=1$ in  (\ref{eq:H_RWA}) and (\ref{eq:H_MS}). Based on the interaction part of these Hamiltonians, we define the effective field operators that drive the dynamics of the field-atom system. Dependent on $\omega$ and the harmonic indices, they read
\begin{equation}
 \op{a}_{\bar{n}}(\omega)=\frac{1}{\kappa_{\bar{n}}(\omega)}\frac{\omega^2}{c^2}\int\mathrm{d^3}r \sqrt{\frac{\epsilon^{\prime\prime}(\mathbf{r},\omega)}{\hbar\pi\epsilon_0}}\mathbf{d}\cdot\bar{\bar{G}}^{(\bar{n})}(\mathbf{r_1},\mathbf{r},\omega)\cdot\op{\mathbf{f}}(\mathbf{r},\omega),\label{eq:cont_eff_field_op}
\end{equation}
with $\bar{n}=(n,m,p)$. Definition (\ref{eq:f_op_exp}) along with the orthogonality relations of the spherical vector harmonics ensure that $\op{a}_{\bar{n}}(\omega)$ indeed annihilates excitations associated with the harmonic term and parity $(n,m,p)$:
\begin{eqnarray}
 \begin{aligned}
 \op{a}_{\bar{n}}(\omega)=&\!\!\sum_{K=M,N,L}\int_0^\infty\!\!\!\!\!\mathrm{d}q\left\{\frac{1}{\kappa_{\bar{n}}(\omega)}\frac{\omega^2}{c^2}\int\!\!\mathrm{d^3}r\sqrt{\frac{\epsilon^{\prime\prime}(\mathbf{r},\omega)}{\hbar\pi\epsilon_0\,\, Q_{\bar{n}}^K(q)}}\right.\\
  \times& \mathbf{d}\cdot\bar{\bar{G}}^{(\bar{n})}(\mathbf{r_1},\mathbf{r},\omega)\cdot\mathbf{K}_{\bar{n}}(\mathbf{r},q)\Bigg\}\op{F}^K_{\bar{n}}(q,\omega),\label{eq:cont_eff_field_op2}
 \end{aligned}
\end{eqnarray}
where $\kappa_{\bar{n}}(\omega)$ is the atom-field coupling (yet to be determined) and $Q^K_{\bar{n}}(q)\equiv Q^K_{nm}(q)$. In terms of definition (\ref{eq:cont_eff_field_op2}), the interaction Hamiltonian assumes the form
\begin{equation}
 \op{H}_{int}=-\mathrm{i}\hbar\sum_{\bar{n}}\!\int_0^\infty\!\!\!\!\!\!\!\mathrm{d}\omega\left[\kappa_{\bar{n}}(\omega)\op{a}_{\bar{n}}(\omega)\op{\sigma}_{+}-\mathrm{H.c.}\right].\label{eq:H_int_eff0}
\end{equation}
To determine $\kappa_{\bar{n}}(\omega)$, we require that the excitations created by $\op{a}^\dagger_{\bar{n}}(\omega)$ be normalized, thus
\begin{equation}
 \left[\op{a}_{\bar{n}}(\omega),\op{a}^\dagger_{\bar{n}^\prime}(\omega^\prime)\right]\equiv\delta_{\bar{n}\bar{n}^\prime}\delta(\omega-\omega^\prime).\label{eq:comm_rel_a}
\end{equation}
Substituting the definition of the effective operators and using the commutation relations (\ref{eq:comm_rel_f}), as well as the Green tensor identity for non-magnetic materials (see \cite{Vogel2006})
\begin{eqnarray}
\begin{aligned}
 \int\!\!\mathrm{d^3}r\frac{\omega^2}{c^2}\epsilon^{\prime\prime}(\mathbf{r},\omega)\bar{\bar{G}}^{(\bar{n})}(\mathbf{r_1},\mathbf{r},\omega)\cdot {\bar{\bar{G}}^{(\bar{n})}}^\dagger(\mathbf{r_2},\mathbf{r},\omega)\\
 =\mathfrak{Im}\left[\bar{\bar{G}}^{(\bar{n})}(\mathbf{r_1},\mathbf{r_2},\omega)\right],
 \end{aligned}
\end{eqnarray}
for the atom-field coupling we obtain
\begin{equation}
 |\kappa_{\bar{n}}(\omega)|^2=\frac{1}{\hbar\pi\epsilon_0}\frac{\omega^2}{c^2}\mathbf{d}\cdot\mathfrak{Im}\left[\bar{\bar{G}}^{(\bar{n})}(\mathbf{r_1},\mathbf{r_1},\omega)\right]\cdot\mathbf{d}^*. \label{eq:atom-field_coupling1}
\end{equation}
In order to express $\op{H}_F$ in terms of the effective field operators, we make the following consideration. Since $\op{a}_{\bar{n}}(\omega)$ is a particular linear combination of operators $\op{F}^{M,N,L}_{\bar{n}}(q,\omega)$ (as well as $\op{f}_k(\mathbf{r},\omega)$), the field-atom dynamics driven by it involves only a certain subspace of the total Hilbert space. States belonging to the rest of the Hilbert space will be decoupled from the dynamics. Thus, if we manage to separate the original Hilbert space into orthogonal "bright" and "dark" subspaces, we can construct an effective model by keeping the first and ignoring the latter.

With a procedure, somewhat similar to the Gram-Schmidt orthogonalization, we construct the set of the dark field operators. In the following, we will use the condensed definition
\begin{equation}
 \op{a}_{\bar{n}}(\omega)=\sum_K\int_0^\infty\!\!\!\!\!\mathrm{d}q\,\, \alpha_{\bar{n}}^K(\omega,q)\op{F}^K_{\bar{n}}(q,\omega),\label{eq:cont_eff_field_op3}
\end{equation}
which is identical with (\ref{eq:cont_eff_field_op2}). A set of operators can now be defined in a way that the states excited by them will always be orthogonal to the states excited by the "bright" operators $ \op{a}_{\bar{n}}(\omega)$:
\begin{eqnarray}
 \begin{aligned}
 \op{D}^K_{\bar{n}}(q,\omega)=&\op{F}^K_{\bar{n}}(q,\omega)+\int_0^\infty\!\!\!\!\!\mathrm{d}\omega^\prime\left[\op{a}^\dagger_{\bar{n}}(\omega^\prime),\op{F}^K_{\bar{n}}(q,\omega)\right]\op{a}_{\bar{n}}(\omega^\prime)\\
 =&\op{F}^K_{\bar{n}}(q,\omega)-\alpha_{\bar{n}}^{K*}(\omega,q) \op{a}_{\bar{n}}(\omega).\label{eq:dark_op}
 \end{aligned}
\end{eqnarray}
The orthogonality is easy to test by simply taking the commutator between the dark and bright operators and finding that it is always zero. As a final step, we use definitions (\ref{eq:cont_eff_field_op3}), (\ref{eq:dark_op}), as well as the property (following from (\ref{eq:comm_rel_a}))
\begin{equation}
 \sum_{K}\int_0^\infty\!\!\!\!\!\mathrm{d}q\,\,\alpha_{\bar{n}}^K(\omega,q)\alpha_{\bar{n}^\prime}^{K*}(\omega,q)=\delta_{\bar{n}\bar{n}^\prime},
\end{equation}
we obtain
\begin{eqnarray}
 \begin{aligned}
 \sum_{K}&\!\!\int_0^\infty\!\!\!\!\!\mathrm{d}q\,\,\op{F}^{K\dagger}_{\bar{n}}(q,\omega)\op{F}^K_{\bar{n}}(q,\omega)\\
 &=\op{a}^\dagger_{\bar{n}}(\omega)\op{a}_{\bar{n}}(\omega)+\sum_{K}\!\!\int_0^\infty\!\!\!\!\!\mathrm{d}q\,\,\op{D}^{K\dagger}_{\bar{n}}(q,\omega)\op{D}^K_{\bar{n}}(q,\omega).
 \end{aligned}
\end{eqnarray}
Thus, the non-interacting field Hamiltonian reads
\begin{equation}
 \op{H}_F=\!\!\int_0^\infty\!\!\!\!\!\!\!\mathrm{d}\omega \hbar \omega\! \sum_{\bar{n}}\!\left(\op{a}_{\bar	{n}}^\dagger(\omega)\op{a}_{\bar{n}}(\omega)\! +\!\sum_K\!\int_0^\infty\!\!\!\!\!\!\!\mathrm{d}q\op{D}_{\bar{n}}^{K\dagger}(q,\omega)\op{D}_{\bar{n}}^K(q,\omega)\right).
\end{equation}
Since the operator (\ref{eq:cont_eff_field_op}) doesn't couple to the dark modes, i.e., the states created by the dark operators (\ref{eq:dark_op}), these will have an independent dynamics and will not affect the dynamics of the bright modes.

We can thus work with the effective Hamiltonian restricted to the bright subspace, given as
\begin{eqnarray}
 \begin{aligned}
 \op{H}^{eff}=&\int_0^\infty\!\!\!\!\mathrm{d}\omega\hbar\omega\sum_{\bar{n}}\op{a}^\dagger_{\bar{n}}(\omega)\op{a}_{\bar{n}}(\omega)+\hbar\omega_A\op{\sigma}_{ee}\\
 -&\mathrm{i}\hbar\!\!\int_0^\infty\!\!\!\!\mathrm{d}\omega\sum_{\bar{n}}\left[\kappa_{\bar{n}}(\omega)\op{a}_{\bar{n}}(\omega)\op{\sigma}_+-\mathrm{H.c.}\right],\label{eq:H_eff_single}
 \end{aligned}
\end{eqnarray}
where a single, two-level quantum emitter interacts with the mode structure of the environment. Thus, by constructing the continuous effective model in case of a single emitter, the full, 3D model is projected on an effective 1D cavity-like system (the resonance structure defined by $\kappa_{\bar{n}}(\omega)$). Note that this model is applicable only if the initial state of the field is separable into a bright and a dark component, i.e., there is no initial dark-bright entanglement.

\subsubsection*{Eliminating indices}

The previously described method also enables us to eliminate some of the harmonic indices in case they are not relevant to the current investigation. Let us take an example where we are only interested in the radial harmonic index, $n$, instead of the full $\bar{n}=(n,m,p)$. In this case, it is desirable to construct an effective model where $n$ is the only index present. Taking (\ref{eq:H_int_eff0}), it is obvious that by defining the field operator
\begin{equation}
 \op{a}_n(\omega)=\frac{1}{\kappa_n(\omega)}\sum_{m=0}^n\sum_{p=e,o}\kappa_{\bar{n}}(\omega)\op{a}_{\bar{n}}(\omega),
\end{equation}
we can express the interaction Hamiltonian as
\begin{equation}
 \op{H}_{int}=-\mathrm{i}\hbar\sum_{n=0}^\infty\!\int_0^\infty\!\!\!\!\!\!\!\mathrm{d}\omega\left[\kappa_{n}(\omega)\op{a}_{n}(\omega)\op{\sigma}_{+}-\mathrm{H.c.}\right].\label{eq:H_int_eff_reduced}
\end{equation}
The value of the atom-field coupling $\kappa_n(\omega)$ is, again, defined by the normalization of the field operator, i.e.,
\begin{equation}
 \left[\op{a}_{n}(\omega),\op{a}^\dagger_{n^\prime}(\omega^\prime)\right]\equiv\delta_{nn^\prime}\delta(\omega-\omega^\prime),\label{eq:comm_rel_a_reduced}
\end{equation}
so that we obtain
\begin{eqnarray}
\begin{aligned}
 |\kappa_n(\omega)|^2&=\sum_{m,p}|\kappa_{\bar{n}}(\omega)|^2\\
 &=\frac{1}{\hbar\pi\epsilon_0}\frac{\omega^2}{c^2}\mathbf{d}\cdot\mathfrak{Im}\left[\bar{\bar{G}}^{(n)}(\mathbf{r_1},\mathbf{r_1},\omega)\right]\cdot\mathbf{d^*}.
 \end{aligned}
\end{eqnarray}
To construct the full effective Hamiltonian, we perform the dark-bright subspace separation, defining dark operators
\begin{eqnarray}
 \begin{aligned}
\op{d}_{\bar{n}}(\omega)=\op{a}_{\bar{n}}(\omega)+\int_0^\infty\!\!\!\!\!\mathrm{d}\omega^\prime\left[\op{a}^\dagger_n(\omega^\prime),\op{a}_{\bar{n}}(\omega)\right]\op{a}_n(\omega^\prime).
 \end{aligned}
\end{eqnarray}
Based on this, the dark-bright separation yields the effective Hamiltonian
\begin{eqnarray}
 \begin{aligned}
 \op{H}^{eff}=&\int_0^\infty\!\!\!\!\mathrm{d}\omega\hbar\omega\sum_{n=0}^\infty\op{a}^\dagger_{n}(\omega)\op{a}_{n}(\omega)+\hbar\omega_A\op{\sigma}_{ee}\\
 -&\mathrm{i}\hbar\!\!\int_0^\infty\!\!\!\!\mathrm{d}\omega\sum_{n=0}^\infty\left[\kappa_{n}(\omega)\op{a}_{n}(\omega)\op{\sigma}_+-\mathrm{H.c.}\right].\label{eq:H_eff_single_elim}
 \end{aligned}
\end{eqnarray}
It is very important to note that the dark-bright separation is feasible only if the spectrum of the non-interacting Hamiltonian (in our case $\hbar\omega$) does not depend on the eliminated variables. Thus, for example, we could construct effective models by eliminating the variable $q$ or additional indices because $\hbar\omega$ does not depend either on $q$, or on any of $n$, $m$, or $p$.

\subsection{Continuous effective model with multiple two-level emitters}\label{sec:effective_cont_multiple}

\subsubsection{Several emitters and mode overlap}

Let us now consider the case of $N_e$ two-level emitters interacting with the mode structure characterized by $\bar{\bar{G}}(\mathbf{r},\mathbf{r^\prime},\omega)$. Corresponding to (\ref{eq:H_RWA}) and (\ref{eq:H_MS}), one can define the effective interaction Hamiltonian
\begin{equation}
  \op{H}_{int}=-\mathrm{i}\hbar\sum_{j=1}^{N_e}\sum_{\bar{n}}\!\int_0^\infty\!\!\!\!\!\!\!\mathrm{d}\omega\left[\kappa^j_{\bar{n}}(\omega)\op{a}^j_{\bar{n}}(\omega)\op{\sigma}^j_{+}-\mathrm{H.c.}\right],\label{eq:H_int_eff}
\end{equation}
where now we have a set of creation and annihilation operators for each atomic position $\mathbf{r}_j$:
\begin{eqnarray}
 \begin{aligned}
 \op{a}^j_{\bar{n}}(\omega)=&\!\!\sum_{K=M,N,L}\int_0^\infty\!\!\!\!\!\mathrm{d}q\left\{\frac{1}{\kappa^j_{\bar{n}}(\omega)}\frac{\omega^2}{c^2}\int\!\!\mathrm{d^3}r\sqrt{\frac{\epsilon^{\prime\prime}(\mathbf{r},\omega)}{\hbar\pi\epsilon_0 \,\,Q_{\bar{n}}^K(q)}}\right.\\
  \times& \mathbf{d^j}\cdot\bar{\bar{G}}^{(\bar{n})}(\mathbf{r}_j,\mathbf{r},\omega)\cdot\mathbf{K}_{\bar{n}}(\mathbf{r},q)\Bigg\}\op{F}^K_{\bar{n}}(q,\omega),\label{eq:cont_eff_field_op4}
 \end{aligned}
\end{eqnarray}
analogously with the single-emitter case, and the atom-field coupling $\kappa^j_{\bar{n}}(\omega)$ is defined as (\ref{eq:atom-field_coupling1}) for the respective emitter positions $\mathbf{r}_j$. 

Note that the basis of spherical vector harmonics requires parameters $n,m,p$ and $q$ to describe an arbitrary field distribution. Constructing the effective field operators (\ref{eq:cont_eff_field_op4}), we integrated over the variable $q$. Thus, with only $\bar{n}=(nmp)$, the basis ceases to be complete. That is why, unlike for the full mode-selective quantization (\ref{eq:H_MS}), it is no longer possible to describe the field with a single set of operators, but we have  $\{\op{a}_{\bar{n}}^j(\omega)\}_{j=1}^{N_e}$. Thus, instead of $N_e$ emitters coupled to the mode structure of a \emph{single}, 3D sphere, we get the interaction Hamiltonian of $N_e$ 1D cavities which have a \emph{single} emitter inside them each (\ref{eq:H_int_eff}).

\begin{figure}
 \includegraphics[width=0.4\textwidth]{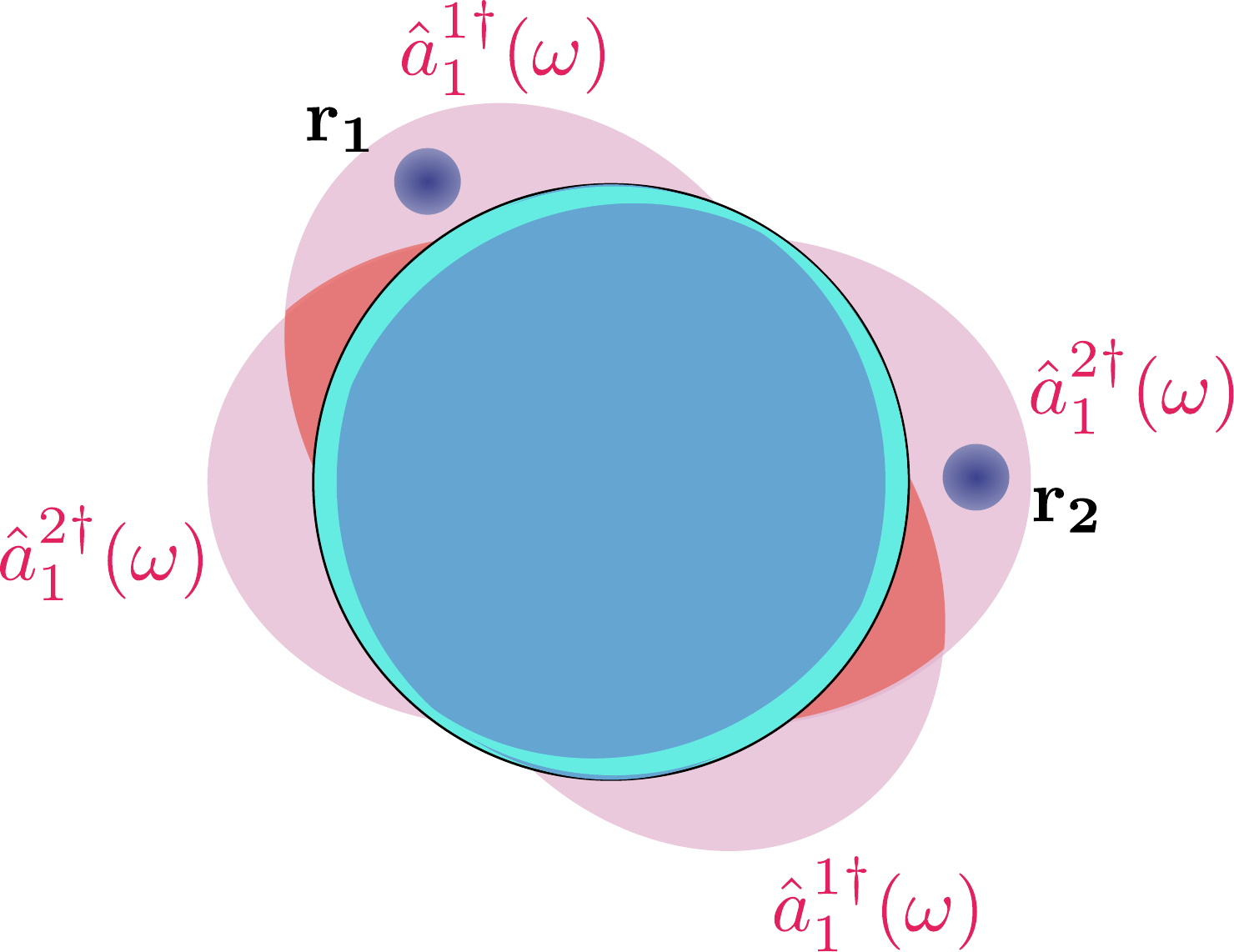}
 \caption{In case of several emitters, field operators are assigned to each emitter position, breaking the angular degeneracy of the harmonic modes. The Figure illustrates the dipolar ($n=1$) modes of a single metallic sphere excited by emitters at $\mathbf{r_1}$ and $\mathbf{r_2}$. Although each emitter is directly coupled only to the field assigned to its position, the overlap of the modes (represented by the red region) can still induce a transfer of excitation between them - hence the non-orthogonality of the field operators.  \label{fig:mode_overlap}}
\end{figure}

The commutation relations for the effective field operators can be obtained with some work. Keeping in mind that, due to the orthogonality of the vector harmonics ($\mathbf{K}=\mathbf{M},\mathbf{N},\mathbf{L}$),
\begin{eqnarray}
\begin{aligned}
 \sum_{l=1}^3 &\int\!\!\mathrm{d^3}r^\prime G^{(\bar{n})}_{kl}(\mathbf{r},\mathbf{r^\prime}) K_{\bar{n}l}(\mathbf{r^\prime},q)\\
 &=\sum_{l=1}^3 \int\!\!\mathrm{d^3}r^\prime G^{(\bar{n})}_{kl}(\mathbf{r},\mathbf{r^\prime})\sum_{\bar{n}^\prime} K_{\bar{n}^\prime l}(\mathbf{r^\prime},q),
 \end{aligned}
\end{eqnarray}
we take (\ref{eq:cont_eff_field_op4}), as well as relations (\ref{eq:comm_rel_F1}) and (\ref{eq:comm_rel_F2}). Simplifying the resulting expression by  aid of the Dirac delta expansion in terms of spherical harmonics (see Appendix \ref{sec:app_Dirac}) it becomes apparent that the creation and annihilation operators cease to be orthogonal for different emitter positions:
  \begin{equation}
 \left[\op{a}^i_{\bar{n}}(\omega),\op{a}^{j\dagger}_{\bar{n}^\prime}(\omega^\prime)\right]\equiv\delta_{\bar{n}\bar{n}^\prime}\delta(\omega-\omega^\prime)\mu^{ij}_{\bar{n}}(\omega),\label{eq:comm_rel_a2}
\end{equation}
where we call $\mu^{ij}_{\bar{n}}(\omega)$ the \emph{mode overlap} and it reads
\begin{equation}
 \mu_{\bar{n}}^{ij}(\omega)=\frac{1}{\hbar\pi\epsilon_0}\frac{\omega^2}{c^2}\frac{\mathbf{d}^i\cdot\mathfrak{Im}\left[\bar{\bar{G}}^{(\bar{n})}(\mathbf{r}_i,\mathbf{r}_j,\omega)\right]\cdot\mathbf{d}^{j*}}{\kappa^i_{\bar{n}}(\omega)\kappa^{j*}_{\bar{n}}(\omega)}.\label{eq:mode_overlap}
\end{equation}
The non-orthogonality of the field operators and the mode overlap is illustrated in Fig. \ref{fig:mode_overlap}. With (\ref{eq:atom-field_coupling1}), it is easily seen that $\mu^{jj}_{\bar{n}}(\omega)=1$, thus the $N_e$-emitter model returns the single-emitter commutation relations for $N_e=1$. Following the treatment seen previously, the dark-bright subspace separation is supposed to lead to the desired effective Hamiltonian. 
%

However, the presence of the non-unity mode overlap poses a problem. Since, for different $j$ parameters, the bright operators $\op{a}^j_{\bar{n}}(\omega)$ are not necessarily orthogonal, trying to construct the dark operators similarly to (\ref{eq:dark_op}) will not result in a set of operators that commute with the bright ones. That is to say, the Hilbert space will no longer separate into two, orthogonal subspaces if we follow this procedure. 
  
Instead, one must include an intermediate step: the orthogonalization of the bright operator manifold with respect to the emitter position parameter. As a first step, as seen below, we reduce the manifold of the original bright operators to linearly independent ones.

\subsubsection{Non-independent field operators}

 In a given arrangement, it can happen that the mode overlap for two emitters is unity. If one aims for maximizing the plasmon-mediated coupling between two emitters, this is even a desirable situation. However, in such a case the  bright operators belonging to each of the atoms do not create linearly independent plasmonic states - and this poses a problem if we want to orthogonalize the manifold of the bright operators. We can handle this issue by expressing two non-independent field operators with a single one, ultimately reducing the manifold to linearly independent operators.
 
 If, for a given pair of atomic position indices $(i,j)$, where $i>j$, we have
\begin{equation}
 |\mu_{\bar{n}}^{ij}(\omega)|=1,\label{eq:lin_dependence}
\end{equation}
then it means that $\op{a}^i_{\bar{n}}(\omega)$ and $\op{a}^j_{\bar{n}}(\omega)$ overlap completely and, as a consequence, are not linearly independent. This becomes apparent if we multiply both sides of (\ref{eq:comm_rel_a2}) with ${\mu^{ij}_{\bar{n}^\prime}}^*(\omega^\prime)$. If (\ref{eq:lin_dependence}) is true, we have
\begin{equation}
 \left[\op{a}^i_{\bar{n}}(\omega),\left(\mu^{ij}_{\bar{n}^\prime}(\omega^\prime)\op{a}^j_{\bar{n}^\prime}(\omega^\prime)\right)^\dagger\right]=\delta_{\bar{n}\bar{n}^\prime}\delta(\omega-\omega^\prime),\label{eq:comm_rel_a3}
\end{equation}
and, as a consequence, 
\begin{equation}
 \op{a}^i_{\bar{n}}(\omega)=\mu^{ij}_{\bar{n}}(\omega)\op{a}^j_{\bar{n}}(\omega).
\end{equation}

Listing all the index pairs $i>j$ where the absolute value of the mode overlap is $1$ and making the above assignment in each case, we end up with a \emph{reduced} dimensionality $N_r\leq N_e$ for the manifold of the field operators. Also, expressing a pair of completely overlapping field operators with a single one results in the interaction of two atoms with the same field:  
\begin{equation}
\op{H}_{int}=-\mathrm{i}\hbar\sum_{j=1}^{N_r}\sum_{l=1}^{N_e}\sum_{\bar{n}}\!\int_0^\infty\!\!\!\!\!\!\!\mathrm{d}\omega\left[\kappa^{jl}_{\bar{n}}(\omega)\op{a}^j_{\bar{n}}(\omega)\op{\sigma}^l_{+}-\mathrm{H.c.}\right],\label{eq:H_int_eff2}
\end{equation} 
where we have defined the couplings strengths as
\begin{eqnarray}
 \begin{aligned}
 \kappa^{jl}_{\bar{n}}(\omega)&=\left\{\begin{array}{c c c} \mu^{lj}_{\bar{n}}(\omega)\kappa^l_{\bar{n}}(\omega) & \quad\quad & 								       j\leq l\,\, \mathrm{and}\,\,  |\mu^{lj}_{\bar{n}}(\omega)|=1\\
							      0 & \quad & \mathrm{otherwise} \end{array}\right. ,\label{eq:coupling_reduced}
 \end{aligned}
\end{eqnarray}
where 
\begin{equation}
 1 \le j \le N_r \quad \mathrm{and} \quad 1 \le  l \le N_e.
\end{equation}
Note that the above procedure has been applied in \cite{Rousseaux2016} where two emitters are placed symmetrically at opposite sides of a silver sphere in order to maximize their interaction through the plasmons. In the following, we proceed with orthogonalizing this reduced manifold of bright operators.

\subsubsection{Orthogonalizing the bright operators}

Having obtained a manifold of operators which do not overlap completely, we can now orthogonalize them. There are several possible methods to choose from, depending on our convenience. For example, with a Gram-Schmidt type of orthogonalization, one can construct
\begin{eqnarray}
 \begin{aligned}
  \op{b}^1_{\bar{n}}(\omega)&=\op{a}^1_{\bar{n}}(\omega)\\
  \beta_2\,\op{b}^2_{\bar{n}}(\omega)&=\left(\op{a}^2_{\bar{n}}(\omega)+\int_0^\infty\!\!\!\!\!\mathrm{d}\omega^\prime\left[\op{b}^{1\dagger}_{\bar{n}}(\omega^\prime),\op{a}^2_{\bar{n}}(\omega)\right]\op{b}^1_{\bar{n}}(\omega^\prime)\right)\\
  &\vdots\\
  \beta_{N_r}\op{b}^{N_r}_{\bar{n}}(\omega)\!\!&= \\
  \times&\bigg(\op{a}^{N_r}_{\bar{n}}(\omega)\!+\!\!\int_0^\infty\!\!\!\!\!\mathrm{d}\omega^\prime\!\!\sum_{j=1}^{N_r-1}\!\left[\op{b}^{j\dagger}_{\bar{n}}(\omega^\prime),\op{a}^{N_{red}}_{\bar{n}}(\omega)\right]\op{b}^j_{\bar{n}}(\omega^\prime)\bigg),\\ \label{eq:orth_modes_G-S}
 \end{aligned}
\end{eqnarray}
where the coefficients $\beta_j$ ensure normalization. Depending on our needs, we can choose other methods: for example, the Householder reflection makes the orthogonalized basis inherit the symmetries of the original  one. In general, we can represent the manifold of the orthogonal bright operators as
\begin{equation}
 \op{b}^i_{\bar{n}}(\omega)=\sum_{j=1}^{N_r} B^{\bar{n}}_{ij}(\omega)\op{a}^j_{\bar{n}}(\omega),\label{eq:b_expansion_a}
\end{equation}
where, since the operators $\op{a}^j_{\bar{n}}(\omega)$ are linearly independent, the matrix made up by the elements $B^{\bar{n}}_{ij}(\omega)$ is non-singular. Taking the inverse of (\ref{eq:b_expansion_a}), we can write
\begin{equation}
 \op{a}^j_{\bar{n}}(\omega)=\sum_{i=1}^{N_r} A^{\bar{n}}_{ji}(\omega)\op{b}^i_{\bar{n}}(\omega).\label{eq:a_expansion_b}
\end{equation}

\begin{figure}
 \begin{center}
  \includegraphics[width=0.4\textwidth,angle=0]{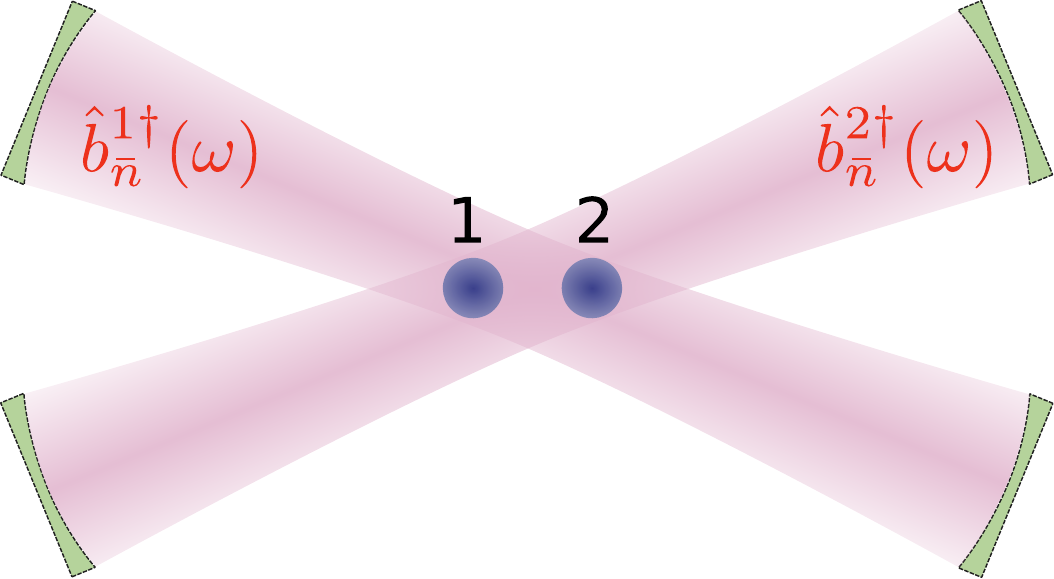}
  \caption{Illustration of the effective model with commuting bright operators $\{\hat{b}^j_{\bar{n}}(\omega)\}_{j=1}^{N_r}$, obtained from the orthogonalization of operators $\{\hat{a}^j_{\bar{n}}(\omega)\}_{j=1}^{N_r}$, for an example of two emitters where $N_e=N_r=2$. It can be represented as two 1D cavities where each emitter may interact with both cavity fields. \label{fig:mode_overlap_cavities}}
 \end{center}
\end{figure}

Thanks to the orthonormality of the set, we have now
\begin{equation}
 \left[\op{b}^i_{\bar{n}}(\omega),\op{b}^{j\dagger}_{\bar{n}^\prime}(\omega^\prime)\right]\equiv\delta_{\bar{n}\bar{n}^\prime}\delta(\omega-\omega^\prime)\delta_{ij},\label{eq:comm_rel_b}
\end{equation}
and thus we can perform the dark-bright separation as described in the previous Section - only instead of the original bright operators we use the reduced, orthonormalized set $\{\op{b}^i_{\bar{n}}(\omega)\}$. Applying the steps described therein, we arrive at the effective Hamiltonian for $N_e$ emitters interacting with the surrounding mode structure:
\begin{eqnarray}
 \begin{aligned}
 \op{H}^{eff}=&\int_0^\infty\!\!\!\!\mathrm{d}\omega\hbar\omega\sum_{i=1}^{N_r}\sum_{\bar{n}}\op{b}^{i\dagger}_{\bar{n}}(\omega)\op{b}^i_{\bar{n}}(\omega)+\sum_{l=1}^{N_e}\hbar\omega_A\op{\sigma}^l_{ee}\\
 -&\mathrm{i}\hbar\!\!\int_0^\infty\!\!\!\!\mathrm{d}\omega\sum_{i=1}^{N_r}\sum_{l=1}^{N_e}\sum_{\bar{n}}\left[\tilde{\kappa}^{il}_{\bar{n}}(\omega)\op{b}^i_{\bar{n}}(\omega)\op{\sigma}^l_+-\mathrm{H.c.}\right],\label{eq:H_eff_multi}
 \end{aligned}
\end{eqnarray}
where now each atom interacts with all the collective fields, with atom-field couplings
\begin{equation}
 \tilde{\kappa}^{il}_{\bar{n}}(\omega)=\sum_{j=1}^{N_r}A_{ji}^{\bar{n}}(\omega)\kappa^{jl}_{\bar{n}}(\omega).\label{eq:coupling_multi_emitter}
\end{equation}
Thus, the final effective model is an ensemble of $N_r$ 1D cavity-like systems (each with its own set of operators $\op{b}^j_{\bar{n}}(\omega)$) where now the fields of the empty cavities commute (there is no mode overlap), but \emph{each} emitter may interact with \emph{all} cavity fields. A visual representation is given in Fig. \ref{fig:mode_overlap_cavities}.

\subsubsection{Two-emitter systems: two examples}

To consider a particular implementation of the treatment described above, we will now look at a system where two quantum emitters are in the close vicinity of a nanosphere, thus, $N_e=2$. A possible arrangement is represented in Fig. \ref{fig:mode_overlap}. As relevant mode index we take $n$ so the interaction Hamiltonian with the initial bright operators reads
\begin{eqnarray}
 \begin{aligned}
  \op{H}_{int}=-\mathrm{i}\hbar\sum_{n}\!\int_0^\infty\!\!\!\!\!\!\!\mathrm{d}\omega&\left[\kappa^1_{n}(\omega)\op{a}^1_{n}(\omega)\op{\sigma}^1_{+}+\kappa^2_{n}(\omega)\op{a}^2_{n}(\omega)\op{\sigma}^2_{+}\right.\\
  &\left.-\mathrm{H.c.}\right],
 \end{aligned}
\end{eqnarray}
where the couplings $\kappa^{1,2}_{n}(\omega)$ are defined by (\ref{eq:atom-field_coupling1}) for the respective emitter positions $\mathbf{r_{1,2}}$. The mode overlap enters with the commutation relation
\begin{equation}
 \left[\op{a}^2_{n}(\omega),\op{a}^{1\dagger}_{n^\prime}(\omega^\prime)\right]\equiv\delta_{nn^\prime}\delta(\omega-\omega^\prime)\mu^{21}_{n}(\omega).
\end{equation}
We first consider a general arrangement where the two modes are only partially overlapping, that is, $|\mu^{21}_{n}(\omega)|\neq 1$. Thus, the coupling strengths defined according to (\ref{eq:coupling_reduced}) read
\begin{equation}
 \kappa^{ij}_n(\omega)=\kappa^i_n(\omega)\delta_{ij}.
\end{equation}
Having linearly independent modes, we can proceed straight away with diagonalizing the bright operators. Following (\ref{eq:orth_modes_G-S}) as well as requiring commutiation relations (\ref{eq:comm_rel_b}) to apply, we get
\begin{eqnarray}
 \begin{aligned}
  \op{b}_n^1(\omega)&=\op{a}_n^1(\omega)\\
  \op{b}_n^2(\omega)&=\frac{1}{\beta}\left(\op{a}_n^2(\omega)-\mu^{21}_n(\omega)\op{a}_n^1(\omega)\right),\label{eq:b_2_em}
 \end{aligned}
\end{eqnarray}
where
\begin{equation}
 \beta=\sqrt{1-|\mu^{12}_n(\omega)|^2}.
\end{equation}

thus, the matrix of coefficients, defined in (\ref{eq:b_expansion_a}), reads
\begin{equation}
 \bar{\bar{B}}^n(\omega)=\left[\begin{array}{cc}
                          1	&	0\\
                          -\frac{\mu^{21}_n(\omega)}{\beta}	&	\frac{1}{\beta}
                         \end{array}\right].\label{eq:b_expansion_a_2_em}
\end{equation}
Having constructed the new, orthogonal basis of bright operators, we now express the interaction Hamiltonian with them. To do this, we invert (\ref{eq:b_expansion_a_2_em}), getting
\begin{equation}
 \left[\begin{array}{c}
  \op{a}_n^1(\omega)\\
  \op{a}_n^2(\omega)
 \end{array}\right]= \left[\begin{array}{cc}
                            1	&	0\\
                            \mu^{21}_n(\omega)	&	\beta
                           \end{array}\right] \left[\begin{array}{c}
						      \op{b}_n^1(\omega)\\
						      \op{b}_n^2(\omega)
						      \end{array}\right].
\end{equation}
This is all one needs to express the effective Hamiltonian for a two-emitter system where the overlap of the modes excited by the emitters is not complete. Using (\ref{eq:coupling_multi_emitter}), we obtain the couplings to the orthogonalized bright operators and thus 
\begin{eqnarray}
 \begin{aligned}
 \op{H}^{eff}=&\int_0^\infty\!\!\!\!\mathrm{d}\omega\hbar\omega\sum_{i=1}^{2}\sum_{n}\op{b}^{i\dagger}_{n}(\omega)\op{b}^i_{n}(\omega)+\sum_{l=1}^{2}\hbar\omega_A\op{\sigma}^l_{ee}\\
 -&\mathrm{i}\hbar\!\!\int_0^\infty\!\!\!\!\mathrm{d}\omega\sum_{n}\left[\kappa^1_n(\omega)\op{b}^1_n(\omega)\op{\sigma}^1_+\right.\\ 
 &\left.+\mu^{21}_n\kappa_n^2(\omega)\op{b}^1_n(\omega)\op{\sigma}^2_+ +\beta\kappa^2_n(\omega)\op{b}^2_{n}(\omega)\op{\sigma}^2_+-\mathrm{H.c.}\right].\label{eq:H_eff_multi}
 \end{aligned}
\end{eqnarray}
It is easily seen that atom $1$ and $2$ couple to field $\op{b}^1_n$, while atom $2$ also couples to $\op{b}^2_n$ with a coupling strength that is determined by how imperfect the overlap of the original modes is (i.e., by $\beta$).

In order to demonstrate the case of linearly non-independent modes, we will now consider the arrangement where the emitters are radially polarized and placed at the opposite sides of the sphere, i.e., $\mathbf{r_1}=(r, \pi/2, \phi)$ and $\mathbf{r_2}=(r, \pi/2, \phi+\pi)$. A careful analysis of $\hat{r}_1\cdot\bar{\bar{G}}^{(n)}(\mathbf{r_1}, \mathbf{r_2},\omega)\cdot\hat{r}_2$ in this case yields the result of
\begin{equation}
 \mu^{21}_{n}(\omega) = (-1)^n, 
\end{equation}
thus, with $|\mu^{21}_{n}(\omega)|=1$, we are looking at the situation of fully overlapping modes. Using (\ref{eq:comm_rel_a3}), we can express one of the fields with the other:
\begin{equation}
 \op{a}^2_n(\omega)=(-1)^n\op{a}^1_n(\omega).
\end{equation}
Thus, we have eliminated the second field, obtaining a reduced manifold for the bright operators with $N_r=1$. According to (\ref{eq:coupling_reduced}), we define the matrix of coupling strengths
\begin{equation}
\begin{aligned}
 \kappa^{11}_n(\omega)&=\kappa^1_n(\omega)\\
 \kappa^{12}_n(\omega)&=(-1)^n\kappa^2_n(\omega).
 \end{aligned}
\end{equation}
Having a single field now, there is no need for orthogonalization ($\op{b}^1_n(\omega)=\op{a}^1_n(\omega)$ with $N_r=1$), thus we can write the effective Hamiltonian as
\begin{eqnarray}
 \begin{aligned}
  \op{H}^{eff}=&\int_0^\infty\!\!\!\!\mathrm{d}\omega\hbar\omega\sum_{\bar{n}}\op{b}^{1\dagger}_{n}(\omega)\op{b}^1_{n}(\omega)+\sum_{l=1}^{2}\hbar\omega_A\op{\sigma}^l_{ee}\\
 -&\mathrm{i}\hbar\!\!\int_0^\infty\!\!\!\!\mathrm{d}\omega\sum_{n}\left[\kappa^1_n(\omega)\op{b}^1_n(\omega)\op{\sigma}^1_+ \right.\\ 
 &\left. +(-1)^n\kappa_n^2(\omega)\op{b}^1_n(\omega)\op{\sigma}^2_+-\mathrm{H.c.}\right],\label{eq:H_eff_multi2}
 \end{aligned}
\end{eqnarray}
obtaining the Hamiltonian of a system where the two atoms interact with the same field.

\subsection{Discrete effective model with a single emitter}\label{sec:effective_discrete_single}

So far, we have succeeded in building effective models where the field operators are labelled by the harmonic indices (and the parity $p$) and the sole continuous variable they depend on is $\omega$. In the following derivation, we will get rid of the frequency dependence and construct discrete field operators that are only labelled by the index group $\bar{n}$. Thus, we will have an effective model where each plasmonic resonance peak (in case of a metallic system) has a single, discrete creation/annihilation operator associated to it.

For simplicity, we start with a system where a single quantum emitter interacts with its environment and so the effective Hamiltonian is (\ref{eq:H_eff_single}). Eliminating $\omega$ from the interaction part naturally suggests the standard QED interaction Hamiltonian
\begin{equation}
\op{H}^{eff}_{int}=-\mathrm{i}\hbar\sum_{\bar{n}}\left[g_{\bar{n}}\op{a}_{\bar{n}}\op{\sigma}_{+}-\mathrm{H.c.}\right],
\end{equation}
where the field operator relates to the previously established set as
\begin{equation}
 \op{a}_{\bar{n}}=\frac{1}{g_{\bar{n}}}\int_0^\infty\!\!\!\!\!\mathrm{d}\omega \kappa_{\bar{n}}(\omega)\op{a}_{\bar{n}}(\omega).
\end{equation}
However, the variable that we eliminate is now $\omega$, and it is present in the eigenvalues of the non-interacting Hamiltonian: thus, a dark-bright separation is no longer possible in the way previously discussed. In order to construct a discrete model, we follow a different path. 

Let us write the Schr\"{o}dinger equation for this system, i.e., using the effective Hamiltonian (\ref{eq:H_eff_single}),
\begin{equation}
 \mathrm{i}\hbar\partial_t|\psi(t)\rangle = \op{H}^{eff}|\psi(t)\rangle.
\end{equation}
Having a single emitter, it is reasonable to work in the single-excitation subspace, thus we choose
\begin{equation}
 |\psi(t)\rangle=c_0(t)e^{-\mathrm{i}\omega_A t}|e,0\rangle+\sum_{\bar{n}}\int_0^\infty\!\!\!\!\!\mathrm{d}\omega c_{\bar{n},\omega}(t)e^{-\mathrm{i}\omega t}|g,1_{\bar{n},\omega}\rangle
\end{equation}
for our interaction-picture state vector. $|e,0\rangle$ means that the emitter is in its excited state and the field is in vacuum, while $|g,1_{\bar{n},\omega}\rangle$ represents the emitter being in the ground state and a single excitation in the field, created by $\op{a}_{\bar{n}}(\omega)$. The equation of motion for the probability amplitudes read
\begin{eqnarray}
 \begin{aligned}
\dot{c}_0&=-\sum_{\bar{n}}\!\int_0^\infty\!\!\!\!\!\mathrm{d}\omega\,\kappa_{\bar{n}}(\omega)e^{-\mathrm{i}(\omega-\omega_A)t}c_{\bar{n},\omega}(t)\\
\dot{c}_{\bar{n},\omega}&=\kappa^*_{\bar{n}}(\omega) e^{\mathrm{i}(\omega-\omega_A)t}c_0(t).\label{eq:Schrodinger1}
 \end{aligned}
\end{eqnarray}
In the following we will show that, depending on the spectral dependence of the atom-field couplings, it is possible to regroup  $c_{\bar{n},\omega}$  in order to get probability amplitudes that depend only on the indices $\bar{n}$. 

For simplicity, let us use the model of (\ref{eq:H_eff_single_elim}), thereby having to keep track of the single index $n$ only. Thus, the equations of motion will be the same, except we will have $n=1,2,...$ instead of $\bar{n}=(n,m,p)$. 

In order to proceed, we require an important feature of the atom-field coupling, namely, that it be of a Lorentzian profile: 
\begin{equation}
 \kappa_n(\omega)=g_n \mathcal{L}_n(\omega),\label{eq:coupling_Lorentzian}
\end{equation}
where
\begin{equation}
 \mathcal{L}_n(\omega)=\sqrt{\frac{\gamma_n}{\pi}}\frac{1}{\omega-\omega_n+\mathrm{i}\gamma_n},\label{eq:Lorentzian}
\end{equation}
the parameters $\gamma_n$ and $\omega_n$ being the half-width and the center of the peak, respectively. 

\begin{figure}
 \includegraphics[angle=0,width=0.45\textwidth]{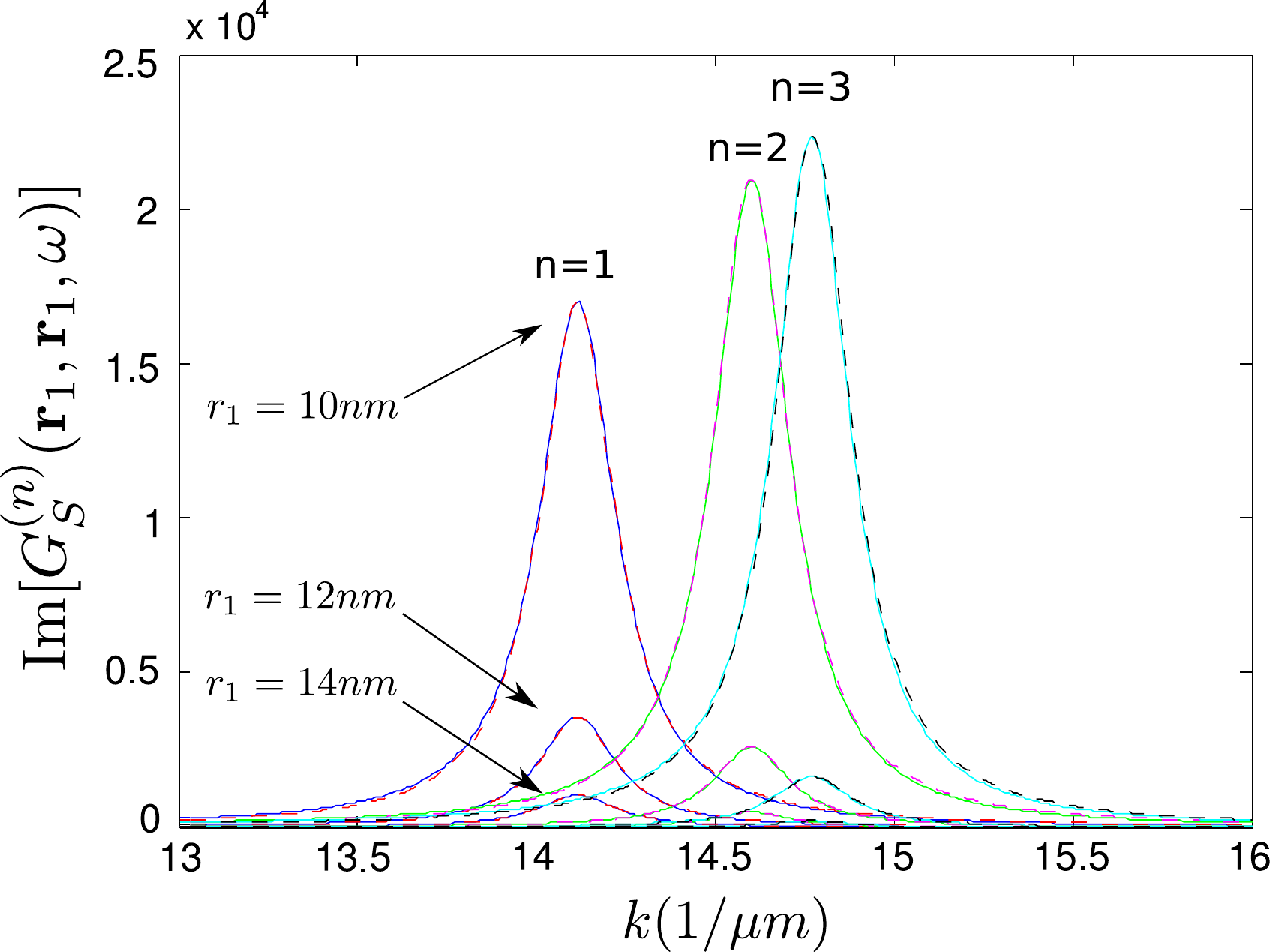}
 \caption{Imaginary part of the $\hat{\mathbf{r}}\otimes \hat{\mathbf{r}}$ scattered Green tensor component induced by a  single silver nanosphere of $8$ nm radius, for different radial coordinates $r_1$ and radial harmonic orders $n$, as a function of the vacuum wave number $k=\omega/c$. The full expression for the scattered Green tensor (continuous lines) and its Lorentzian approximation (dashed lines) show an extremely good correspondence. \label{fig:Lorentz_fit}}
\end{figure}

For a spherical system, examining the structure of the Green tensor in more detail (see Appendix \ref{sec:app_Greens_tensor}), one finds that the resonance-like behaviour of the LDOS is due to the presence of the reflection and transmission coefficients in (\ref{eq:Green_scattered}).  The only other terms with frequency dependence are the radial terms of the spherical vector harmonics: these superimpose oscillations onto the resonance peak. Consequently, if the period of these oscillations is larger than the $\gamma_n$ half-width of the peak (provided by the reflection coefficients), then the Lorentzian lineshape will be a good approximation for the local density of states. Thus we can state that the condition for the applicability of this approximation is
\begin{equation}
 \gamma_n \ll \frac{2\pi c}{r_A\sqrt{\epsilon_f}},
\end{equation}
where $r_A$ is the radial coordinate of the atomic position and $\epsilon_f$ is the relative electric permittivity at the atomic position. Fig. \ref{fig:Lorentz_fit} shows that in case of an emitter interacting with a single metallic sphere, for a sphere and emitter-surface distances of sub-wavelength dimensions this is indeed a very good approximation. 
%
%

Note that there is no need of a numerical fitting procedure in order to determine the center and the half-width of the resonances. Instead, one can construct the so-called mode equation \cite{Derom2012} by requiring a non-trivial solution of the boundary condition equations. The solution of the mode equation is a complex number, the real part and the imaginary part being the central frequency and the half-width of the given resonance, respectively.

\begin{figure}
 \includegraphics[angle=0,width=0.45\textwidth]{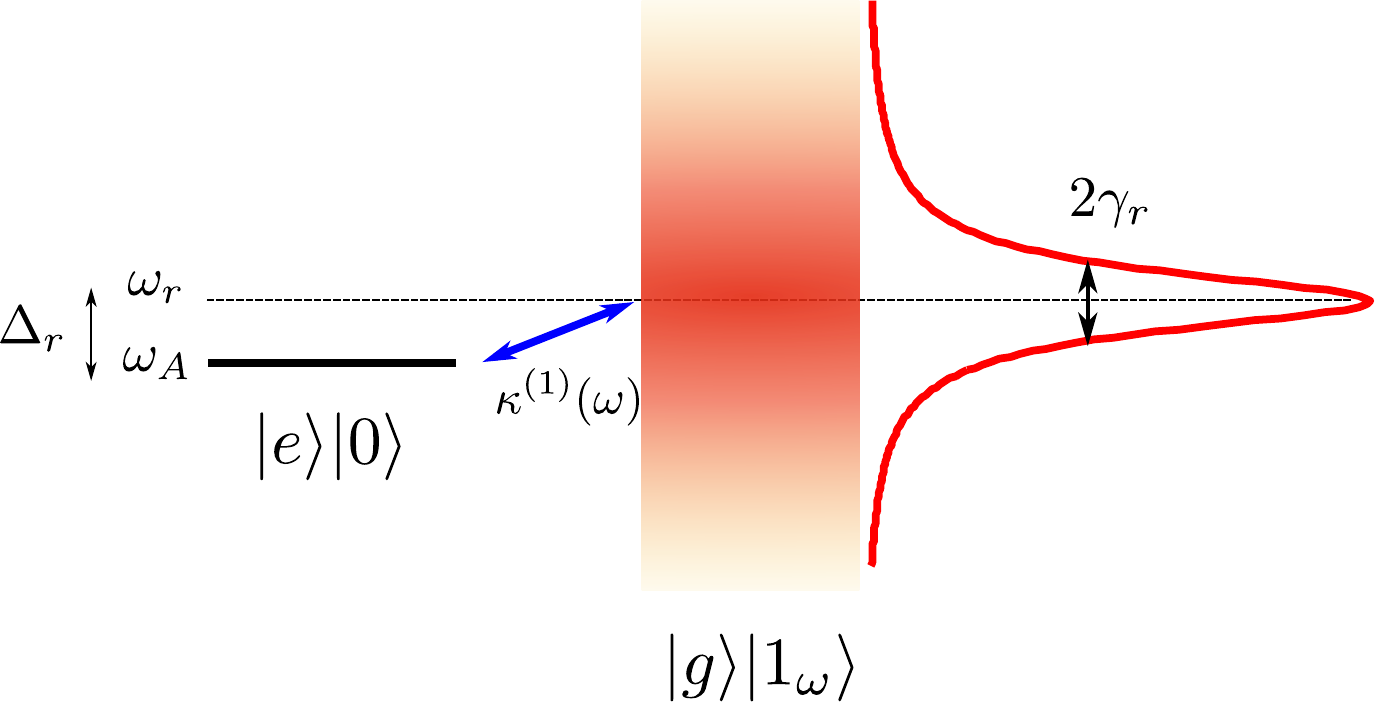}
 \caption{Two-level emitter interacting with a structured continuum, containing a single resonance peak. The dynamics happens in the subspace spanned by the states where either the emitter is excited and the field is in vacuum, or the emitter is in ground state and there is a single excitation in the field.  $\kappa^{(1)}(\omega)$ is the atom-field coupling coefficient. The continuum is structured as a Lorentzian function, with central frequency $\omega_r$ and haf-width $\gamma_r$.   \label{fig:lorentz_illust}}
\end{figure}

Let us define now the collective probability amplitude
\begin{equation}
 c_n(t)=\int_0^\infty\!\!\!\!\!\mathrm{d}\omega \mathcal{L}_n(\omega) e^{-\mathrm{i}(\omega-\omega_A)t}c_{n,\omega}(t).\label{eq:amp_coll}
\end{equation}
Taking the time derivative of it and using the second equation of (\ref{eq:Schrodinger1}), we have
\begin{eqnarray}
 \begin{aligned}
 \dot{c}_n(t)&=g_n^* c_0(t)+\dot{c}_{n,0}(t)-\mathrm{i}g_n^*\!\!\int_{t_0}^t\!\!\!\mathrm{d}t^\prime c_0(t^\prime)\\
 \times&\int_0^\infty\!\!\!\!\!\mathrm{d}\omega (\omega-\omega_A)|\mathcal{L}_n(\omega)|^2 e^{-\mathrm{i}(\omega-\omega_A)(t-t^\prime)},
 \end{aligned}
\end{eqnarray}
where we have defined
\begin{equation}
 c_{n,0}(t)\equiv\int_0^\infty\!\!\!\!\mathrm{d}\omega \mathcal{L}_n(\omega)e^{-\mathrm{i}(\omega-\omega_A)t}c_{n,\omega}(t_0),
\end{equation}
$t_0$ being the instant when the initial conditions were set. Since in a realistic physical situation the contribution given by $\mathcal{L}_n(\omega)$ is negligibly small for $\omega < 0$, we extend the integration to the totality of the real axis. Extending the integration also to the lower complex half plane and using the residue theorem yields
\begin{eqnarray}
 \begin{aligned}
 \dot{c}_n(t)&=g_n^* c_0(t)+\dot{c}_{n,0}(t)\\
 &-\mathrm{i}	g_n^*(\Delta_n-\mathrm{i}\gamma_n)\int_{t_0}^t\!\!\!\mathrm{d}t^\prime e^{-\mathrm{i}(\Delta_n-\mathrm{i}\gamma_n)(t-t^\prime)} c_0(t^\prime),\label{eq:Schrodinger2}
 \end{aligned}
\end{eqnarray}
where $\Delta_n=\omega_n-\omega_A$, i.e., the detuning between the atomic frequency and the center of a given mode. Formally integrating the second equation of (\ref{eq:Schrodinger1}) and substituting it into (\ref{eq:amp_coll}), we arrive to
\begin{equation}
c_n(t)=c_{n,0}(t)+g_n^*\int_{t_0}^t\!\!\!\mathrm{d}t^\prime e^{-\mathrm{i}(\Delta_n-\mathrm{i}\gamma_n)(t-t^\prime)}c_0(t^\prime),
\end{equation}
we can express the integral term and plug it back into (\ref{eq:Schrodinger2}), obtaining
\begin{equation}
\begin{aligned}
 \dot{c}_n(t)&=g_n^* c_0(t)+\dot{c}_{n,0}(t)\\
 &-\mathrm{i}(\Delta_n-\mathrm{i}\gamma_n)\left[c_n(t)-c_{n,0}(t)\right].
 \end{aligned}
\end{equation}
Taking the initial condition where the emitter is excited and the field is in the vacuum state at $t=t_0$, we have $c_{n,0}(t)=\dot{c}_{n,0}(t)=0$, and thus as equations of motion for the probability amplitudes we have
\begin{eqnarray}
 \begin{aligned}
  \dot{c}_0&=-\sum_{n=1}^\infty \!g_n c_n(t)\\
\dot{c}_n&=g_n^* c_0(t)-\mathrm{i}(\Delta_n-\mathrm{i}\gamma_n)c_n(t).\label{eq:Schrodinger3}
 \end{aligned}
\end{eqnarray}
With the above procedure, because of the structured nature of the field continuum, we managed to reduce it to discrete states, thereby getting rid of the continuous frequency dependence. Fig. \ref{fig:lorentz_illust} illustrates the procedure described above with a single resonance peak for the sake of simplicity. Taking into account the first $N$ modes, the Schr\"{o}dinger equation (in interaction picture), written in matrix form reads
\begin{equation}
 \mathrm{i}\hbar\,\,\partial_t\!\left[ \begin{array}{c} c_0 \\ c_1 \\ \vdots \\ c_N \end{array}\right] = \bar{\bar{H}}_N
              \cdot \left[ \begin{array}{c} c_0 \\ c_1 \\ \vdots \\ c_N \end{array}\right],\label{eq:Schrodinger4}
\end{equation}
where 
\begin{equation}
 \bar{\bar{H}}_N =
 \hbar \left[ \begin{array}{c c c c c } 0          & -\mathrm{i}g_1^*    &  -\mathrm{i}g_2^*  &         \cdots         &  -\mathrm{i}g_N^*  \\
                           \mathrm{i} g_1   &  \Delta_1-\mathrm{i}\gamma_1 &            0                &           \cdots       &               0             \\
                           \mathrm{i} g_2   &             0                & \Delta_2-\mathrm{i}\gamma_2 &          \cdots        &               \vdots         \\
                                  \vdots           &         \vdots               &           \cdots            &           \ddots       &               0             \\
                           \mathrm{i} g_N   &             0                &           \cdots             &            0           &\Delta_N-\mathrm{i}\gamma_N   \\
                           \end{array}\right]. \label{eq:H_mx}
\end{equation}
%

Without influencing the dynamics, we can renormalize the zero point of energy by applying the following unitary global phase transformation to the Schr\"{o}dinger equation (\ref{eq:Schrodinger4}):
\begin{equation}
 \bar{\bar{U}}_N=\left[ \begin{array}{c c c} e^{\mathrm{i}\omega_A t} & \cdots   &   0                         \\
                                               \vdots            &  \ddots  &  \vdots                      \\
                                               0                 &   \cdots &    e^{\mathrm{i}\omega_A t}
                  \end{array}\right].
\end{equation}
As a result, we get a discrete effective Hamiltonian which can be written in terms of operators as
\begin{equation}
 \op{H}^{eff}_d\!\!\!=\sum_{n=1}^\infty\hbar(\omega_n-\mathrm{i}\gamma_n)\op{a}_n^\dagger\op{a}_n + \hbar\omega_A \op{\sigma}_{ee}-\mathrm{i}\hbar\sum_{n=1}^\infty\left(g_n\op{a}_n\op{\sigma}_+-\mathrm{H.c.}\right),\label{eq:H_eff_d_single}
\end{equation}
which, along with (\ref{eq:H_mx}), are among the most important results of this paper, having thus derived a model where each field operator $\op{a}_n$ corresponds to a single, lossy harmonic mode that has no continuous dependence on frequency. The parameters of the Lorentzian coupling enter as the peak coupling strength $g_n$, as well as the central frequency $\omega_n$ and spectral half-width $\gamma_n$ of mode $n$. The annihilation operators relate to those of the continuous model as
\begin{equation}
 \op{a}_n=\int_0^\infty\!\!\!\!\!\mathrm{d}\omega\,\mathcal{L}_n(\omega)\op{a}_n(\omega).
\end{equation}
Applying the same procedure for model (\ref{eq:H_eff_single}) yields (\ref{eq:H_eff_d_single}) with the index exchange $n\rightarrow\bar{n}$.

Note that (\ref{eq:H_mx}) can provide a suitable framework for numerical calculations (see \cite{Rousseaux2016} as a good example), containing detunings rather than the original atomic and field energies/frequencies. Also, the structure requirements for the local density of states can be less strict upon deriving the discrete model: resonances with general Fano profiles (as in \cite{Dalton2001} for a lossy, multimode cavity) will also yield a similar Hamiltonian.

\subsection{Discrete effective model with multiple emitters}\label{sec:effective_discrete_multiple}

To construct a discrete effective Hamiltonian in case of several emitters interacting with their environment, we can apply a procedure analogous to the one described previously. However, having had to orthogonalize the original set of modes with respect to emitter positions, the atom-field couplings $\tilde{\kappa}^{jl}_{\bar{n}}(\omega)$ in (\ref{eq:H_eff_multi}) are different from the original, single-index couplings. It is essential to ascertain whether they inherited the Lorentzian resonance profile from the original $\kappa^l_{\bar{n}}(\omega)$ couplings if we want to construct a discrete effective model for multiple emitters. As before, let us consider only a single index, $n$, for the following derivation.

According to definition (\ref{eq:coupling_multi_emitter}), the possibly non-Lorentzian frequency dependence can only come from the coefficients $A_n^{ji}(\omega)$. These, in turn, originate from the construction (\ref{eq:orth_modes_G-S}) and so they contain combinations of the mode overlap $\mu_n^{ij}$, as defined in (\ref{eq:mode_overlap}). 

According to the reasoning in the previous Section, if 
\begin{equation}
 \gamma_n \ll \frac{2\pi c}{\mathrm{max}(\sqrt{\epsilon_j}r_j ,\sqrt{\epsilon_l}r_l)},
\end{equation}
where $\epsilon_j$ and $\epsilon_l$ are the relative electric permittivity values at the location of emitter $j$ and emitter $l$, respectively, then we can make the approximation
\begin{eqnarray}
\begin{aligned}
\mathbf{d}^j \cdot \mathrm{Im}\left[\bar{\bar{G}}^{(n)}(\mathbf{r}_j,\mathbf{r}_l,\omega)\right] \cdot {\mathbf{d}^l}^* &\approx \Omega^{(n)}_{jl}(\mathbf{r}_j,\mathbf{r}_l)\,\, |\mathcal{L}_n(\omega)|^2\\
\mathbf{d}^j \cdot \mathrm{Im}\left[\bar{\bar{G}}^{(n)}(\mathbf{r}_j,\mathbf{r}_j,\omega)\right] \cdot {\mathbf{d}^{j}}^* &\approx \Omega^{(n)}_{j}(\mathbf{r}_j)\,\, |\mathcal{L}_n(\omega)|^2,\\
\end{aligned}
\end{eqnarray}
where $\mathcal{L}_n(\omega)$ is the complex Lorentzian function defined in (\ref{eq:Lorentzian}). Compared to it, the other, position-dependent terms vary so slowly in frequency that they can be regarded as constant over the width of the Lorentzian peak:
\begin{eqnarray}
 \begin{aligned}
  \Omega^{(n)}_{jl}(\mathbf{r}_j,\mathbf{r}_l)&=\frac{\pi}{\gamma_n}\,\,\mathbf{d}^j \cdot \mathrm{Im}\left[\bar{\bar{G}}^{(n)}(\mathbf{r}_j,\mathbf{r}_l,\omega_n)\right] \cdot {\mathbf{d}^l}^*\\
  \Omega^{(n)}_{j}(\mathbf{r}_j)&=\frac{\pi}{\gamma_n}\,\,\mathbf{d}^j \cdot \mathrm{Im}\left[\bar{\bar{G}}^{(n)}(\mathbf{r}_j,\mathbf{r}_j,\omega_n)\right] \cdot {\mathbf{d}^j}^*.\\
 \end{aligned}
\end{eqnarray}

Thus, the mode overlap can be written as
\begin{equation}
 \mu_n^{jl} \approx \frac{\Omega^{(n)}_{jl}(\mathbf{r}_j,\mathbf{r}_l)}{\sqrt{\Omega^{(n)}_{j}(\mathbf{r}_j) {\Omega^{(n)}_{l}}^*(\mathbf{r}_l)}}.
\end{equation}

Since, compared to $\mathcal{L}_n(\omega)$, $\mu_n^{jl}$ can be regarded as constant in frequency, the atom-field couplings between the emitters and the orthogonalized field operators will have the same Lorentzian dependence as the original couplings - only with a modified amplitude. Thus, based on (\ref{eq:coupling_reduced}), (\ref{eq:coupling_multi_emitter}) and (\ref{eq:coupling_Lorentzian}) we can write
\begin{equation}
 \tilde{\kappa}^{il}_n(\omega) \approx \sum_{j=1}^{N_r} A^n_{ji}(\omega_n) g^{jl}_n\mathcal{L}_n(\omega) \equiv \tilde{g}^{il}_n \mathcal{L}_n(\omega)
\end{equation}

Having ascertained this, the same procedure can be applied as by the single-emitter problem. Also, starting from the Hamiltonian (\ref{eq:H_eff_multi}), the derived model is the same, only with the exchange $n \rightarrow \bar{n}=(n,m,p)$. Thus, we obtain the discrete, effective Hamiltonian for N emitters interacting with their environment:
\begin{eqnarray}
 \begin{aligned}
\op{H}^{eff}_d&=\sum_{i=1}^{N_r}\sum_{\bar{n}}\hbar(\omega_{\bar{n}}-\mathrm{i}\gamma_{\bar{n}})\op{b}_{\bar{n}}^{i\dagger}\op{b}_{\bar{n}}^i+\sum_{l=1}^N\hbar\omega_A\op{\sigma}^l_{ee}\\
&-\mathrm{i}\hbar \sum_{i=1}^{N_r}\sum_{l=1}^N\sum_{\bar{n}}\left[\tilde{g}^{il}_{\bar{n}}\,\,\op{b}^i_{\bar{n}}\op{\sigma}^l_+ - \mathrm{H.c.}\right],
 \end{aligned}
\end{eqnarray}
where now we have discrete, lossy modes interacting with the emitters, and the field operators are connected to those with continuous frequency dependence as
\begin{equation}
 \op{b}^i_{\bar{n}}=\int_0^\infty\!\!\!\!\!\mathrm{d}\omega \mathcal{L}_n(\omega) \op{b}^i_{\bar{n}}(\omega).
\end{equation}

\section{Summary}\label{sec:summary}

The methods demonstrated in this paper allow for constructing easy-to-use, simple, effective cavity QED models, based on a full, mode-selective quantization of the system. We have established a clear connection between the resulting atom-field coupling constants and the Green tensor function of the system, through which the geometrical and material properties, as well as the emitter positions enter. 

Two kinds of effective models were presented, namely, those where the field creation and annihilation operators depend on the frequency (continuous models), and those where they are spectrally discrete. For the latter, the structured nature of the local density of states was mandatory, i.e., each of them belong to a resonance peak in the LDOS. 

We have demonstrated the procedure of creating such models through the example of a non-magnetic, spherically multilayered medium. However, the steps of the derivation are applicable in other systems as well, i.e., spheroidally or cylindrically layered, etc. As we have seen in Section \ref{sec:mode-selective_quant}, it is the orthogonality relations of the vector harmonics that allow for a mode-selective quantization in terms of the harmonic indices. In a spherically layered medium, the translational symmetry is broken along one variable ($r$), thus, orthogonality is lost for one of the parameters ($q$). However, since for the discrete harmonic indices ($n,m,p$) the orthogonality is still valid, we were able to construct a mode-selective model in terms of these indices. 
Similarly, when creating a mode-selective model in a layered geometry of a different symmetry, one must pay attention to how the orthogonality relations change due to the layered nature of the system. The resulting model will be mode-selective in terms of those harmonic indices for which the orthogonality relations apply even in the layered medium.

Note that the procedure can potentially be extended to describe interactions between emitters of more complex level structures and the environment surrounding them. In this case, the same field operators will couple to a richer manifold of atomic transition operators.

Thus, this method can be a powerful tool to project a wide variety of situations where quantum emitters interact with a structured reservoir on the effective picture of atoms interacting with the modes of a cavity, being thereby of potential interest in understanding the behaviour of a multitude of systems related to various fields of physics.

\section*{Acknowledgements}

The authors acknowledge financial support from the Agence Nationale de la Recherche: Labex ACTION (ANR-11-LABX-01-
01) and PLACORE (ANR-BS10-0007), COST ACTION MP1403 Nanoscale quantum optics,  from the Conseil Regional de Bourgogne and FEDER (PARI PHOTCOM). Also, support from the Hungarian Academy of Sciences in the frame of the Program of Excellence and the Hungarian-French T\'{E}T\_12\_FR-1-2013-0019 (PHC 30503YD Balaton) Project is gratefully acknowledged.

\section*{Appendix}\label{sec:appendix}

\begin{appendix}

\section{Spherical vector harmonics}\label{sec:app_spherical_harm}

The spherical vector harmonics are eigenvectors of the $\boldsymbol{\nabla}\times\boldsymbol{\nabla}\times$ operator:
\begin{equation}
 \boldsymbol{\nabla}\times\boldsymbol{\nabla}\times\mathbf{F}(\mathbf{r},q)=q^2\mathbf{F}(\mathbf{r},q).\label{eq:app_eigenproblem}
\end{equation}
The $\mathbf{F}$ vector functions can be derived from the spherical scalar solutions of the equation
\begin{equation}
 (\nabla^2+q^2)\psi(\mathbf{r},q)=0,
\end{equation}
which assume the form
\begin{equation}
 \psi_{ nm^e_o}(\mathbf{r},q)=z_n(qr)P^m_n(\cos\theta)\begin{array}{c}
                                                               \cos \\ \sin
                                                              \end{array}(m\phi),
\end{equation}
where $P_n^m(\cos\theta)$ is the associated Legendre polynomials belonging to the $n,m$ spherical orders and, depending on regularization requirements regarding the solution,  $z_n$ can be a spherical Bessel or Hankel function of the first kind. $e$ and $o$ stand for even and odd solutions in $\phi$.

Subsequently, the spherical vector harmonics are constructed as follows:
\begin{eqnarray}
 \begin{aligned}
  \mathbf{M}_{nm^e_o }(\mathbf{r},q)&=\boldsymbol{\nabla}\times\left[\psi_{nm^e_o }(\mathbf{r},q)\mathbf{r}\right]\\
\mathbf{N}_{nm^e_o }(\mathbf{r},q)&=\frac{1}{q}\boldsymbol{\nabla}\times\boldsymbol{\nabla}\times\left[\psi_{nm^e_o }(\mathbf{r},q)\mathbf{r}\right]\\
\mathbf{L}_{nm^e_o }(\mathbf{r},q)&=\boldsymbol{\nabla}\psi_{nm^e_o }(\mathbf{r},q),
 \end{aligned}
\end{eqnarray}
leading to the expressions
\begin{eqnarray}
\begin{aligned}
 \mathbf{M}_{nm^e_o}(\mathbf{r},q)=&\mp\frac{m}{\sin{\theta}}z_n(qr)P_n^m(\cos{\theta})\begin{array}{c}
                                                               \sin \\ \cos \end{array}\!(m\phi)\hat{\theta}\\
                                   -&z_n(qr)\frac{\mathrm{d}P_n^m(\cos{\theta})}{\mathrm{d}\theta}\begin{array}{c}
                                                               \cos \\ \sin \end{array}\!(m\phi)\hat{\phi},\\
\end{aligned}
\end{eqnarray}

\begin{eqnarray}
\mathbf{N}_{nm^e_o}(\mathbf{r},q)&=&\frac{n(n+1)}{qr}z_n(qr)P_n^m(\cos{\theta})\begin{array}{c}
                                                               \cos \\ \sin \end{array}\!\!(m\phi)\hat{r}\nonumber \\
                                   &+&\frac{1}{qr}\frac{\mathrm{d}[r z_n(qr)]}{\mathrm{d}r}\bigg[\frac{\mathrm{d}P_n^m(\cos{\theta})}{\mathrm{d}\theta}\begin{array}{c}
                                                               \cos \\ \sin \end{array}\!\!(m\phi)\hat{\theta}\nonumber \\
                                   &\mp&\frac{m}{\sin{\theta}}P_n^m(\cos\theta)\begin{array}{c}
                                                               \sin \\ \cos \end{array}\!(m\phi)\hat{\phi}\bigg],
\end{eqnarray}

\begin{eqnarray}
\begin{aligned}
\mathbf{L}_{nm^e_o}(\mathbf{r},q)= &\frac{\mathrm{d}z_n(qr)}{\mathrm{d}r}P_n^m(\cos{\theta})\begin{array}{c}
                                                               \cos \\ \sin \end{array}\!\!(m\phi)\hat{r}\\
                                 +&\frac{z_n(qr)}{r}\frac{\mathrm{d}P_n^m(\cos\theta)}{\mathrm{d}\theta}\begin{array}{c}
                                                               \cos \\ \sin \end{array}\!\!(m\phi)\hat{\theta}\\
                                 \mp&\frac{m z_n(qr)}{r\sin{\theta}}P_n^m(\cos{\theta})\begin{array}{c}
                                                               \sin \\ \cos \end{array}\!\!(m\phi)\hat{\phi}.
\end{aligned}
\end{eqnarray}
 
$\mathbf{M}(\mathbf{r},q)$ and $\mathbf{N}(\mathbf{r},q)$ are the continuous-spectrum eigenvectors of (\ref{eq:app_eigenproblem}) belonging to $q^2$ and $\mathbf{L}(\mathbf{r},q)$, which is a curl-less vector function, spans the nullspace of the operator.

The spherical vector harmonics form a complete basis \cite{Tai1993,Chew1995}. 
For readability, we define the notation
\begin{equation}
 \int\!\!\!\mathrm{d^3}r\,\,\mathbf{a}^*(\mathbf{r},q)\cdot\mathbf{b}(\mathbf{r},q^\prime)\equiv\langle \mathbf{a}(q)|\mathbf{b}(q^\prime)\rangle\label{eq:app_scalar_prod}
\end{equation}
with which we define the orthogonality relations below. Taking the solution that is regular around the origin, that is, $z_n(qr)=j_n(qr)$, i.e., the spherical Bessel function of the first kind, we have
\begin{eqnarray}
\begin{aligned}
 \langle\mathbf{\mathbf{K}}_{nmp }(q)&|\mathbf{K^\prime}_{n^\prime m^\prime p^\prime}(q^\prime)\rangle\\
 =&\frac{\pi}{2q^2}\mathcal{Q}^K_{nmp}\,\delta(q-q^\prime)\delta_{KK^\prime}\delta_{nn^\prime}\delta_{mm^\prime}\delta_{pp^\prime}\label{eq:app_ortho2}
 \end{aligned}
\end{eqnarray}
where $\mathbf{K}=\{\mathbf{M},\mathbf{N},\mathbf{L}\}$ are the spherical vector harmonics, and the parity $p$ can be $e$ (even) or $o$ (odd). The normalization factor $\mathcal{Q}^K_{nmp}$ is defined as (\ref{eq:orth_norm_factor}).
\newline 
Note that choosing $z_n=h^{(1)}_n$ leads to the similar orthogonality relations, the only difference being the appearance of a factor of $2$ in the normalization constants.

It is also easily seen that taking only the radial components of $\mathbf{N}$ and $\mathbf{L}$, they too show orthogonality in the spherical harmonic indices. This property comes in handy when expanding the singular part of the direct Green tensor term in the spherical vector harmonic basis.
\newline

\section{Green's tensor in a spherically multilayered medium}\label{sec:app_Greens_tensor}

The Green tensor of a spherically symmetric, multilayered system can be calculated following \cite{Li1994, Tai1993, Chew1995}. Constructing the Green tensor of a layered medium (i.e. one that is homogeneous between the layer interfaces and only changes properties from layer to layer) involves two major steps: an expansion in the tensor-produced basis of the eigenfunctions of (\ref{eq:app_eigenproblem}) for the \emph{homogeneous} medium and the determination of the expansion coefficients by imposing boundary conditions at the layer interfaces.

Generally, the dyadic Green's function of a multilayered medium is written as
\begin{equation}
 \bar{\bar{G}}(\mathbf{r},\mathbf{r^\prime},\omega)=\delta_{fs}\bar{\bar{G}}_0(\mathbf{r},\mathbf{r^\prime},\omega)+\bar{\bar{G}}_S(\mathbf{r},\mathbf{r^\prime},\omega),
\end{equation}
where field point $\mathbf{r}$ and source point $\mathbf{r^\prime}$ are in the layers indexed with $f$ and $s$, respectively. In case $f$ and $s$ are the same, a term appears in the Green tensor that represents direct propagation from $\mathbf{r}$ to $\mathbf{r^\prime}$. $\bar{\bar{G}}_S$ accounts for the propagation between $\mathbf{r}$ and $\mathbf{r^\prime}$ due to the scattering of radiation on the surrounding layers. 

In the basis of spherical vector harmonics, the direct (electric) Green tensor term assumes the form
\begin{eqnarray}
\begin{aligned}
 \bar{\bar{G}}&_0(\mathbf{r},\mathbf{r^\prime},\omega)=\frac{\delta(\mathbf{r}-\mathbf{r^\prime})}{k_s^2}\hat{\mathbf{r}}\otimes\hat{\mathbf{r}}\\
 +&\frac{\mathrm{i}k_s}{4\pi}\sum_{n=0}^\infty\sum_{m=0}^n\sum_{p=e,o}(2-\delta_{m0})\frac{(2n+1)(n-m)!}{n(n+1)(n+m)!}\\
 &\times\left\{\begin{array}{l}
               \mathbf{M}^{(1)}_{nmp}(\mathbf{r},k_s)\otimes\mathbf{M}^{(0)}_{nmp}(\mathbf{r^\prime},k_s)\\
               +\mathbf{N}^{(1)}_{nmp}(\mathbf{r},k_s)\otimes\mathbf{N}^{(0)}_{nmp}(\mathbf{r^\prime},k_s) \quad   r\ge r^\prime\\
               \quad\\
               \mathbf{M}^{(0)}_{nmp}(\mathbf{r},k_s)\otimes\mathbf{M}^{(1)}_{nmp}(\mathbf{r^\prime},k_s)\\
               +\mathbf{N}^{(0)}_{nmp}(\mathbf{r},k_s)\otimes\mathbf{N}^{(1)}_{nmp}(\mathbf{r^\prime},k_s) \quad   r\le r^\prime\\
               \end{array}\right. ,\label{eq:app_G0}
\end{aligned}
 \end{eqnarray}
 where - depending on whether the field or the source point is closer to the origin - one has to choose a spherical Bessel or a spherical Hankel function of the first type (upper indices $(0)$ and $(1)$, respectively) for the radial part of the vector harmonics. This ensures that $\bar{\bar{G}}_0$ is regularized as its spatial arguments tend to the origin or infinity. 

If in an $N$-layered medium $\mathbf{r}$ is located in layer $f$ and $\mathbf{r^\prime}$ in layer $s$, the scattered term reads
\begin{eqnarray}
 &&\bar{\bar{G}}_S(\mathbf{r},\mathbf{r^\prime},\omega)\nonumber\\
 &&=\frac{\mathrm{i}k_s}{4\pi}\sum_{n=0}^\infty\sum_{m=0}^n\sum_{p=e,o}(2-\delta_{m0})\frac{(2n+1)(n-m)!}{n(n+1)(n+m)!}\nonumber\\
 &&\times\left\{(1-\delta_{fN})\mathbf{M}^{(1)}_{nmp}(\mathbf{r},k_f)\otimes\left[(1-\delta_{s1})A_M^{fs}\mathbf{M}^{(0)}_{nmp}(\mathbf{r^\prime},k_s)\right.\right.\nonumber\\
 &&\quad\quad\quad\quad \left.+(1-\delta_{sN})B^{fs}_M\mathbf{M}^{(1)}_{nmp}(\mathbf{r^\prime},k_s)\right]\nonumber\\
 &&+(1-\delta_{fN})\mathbf{N}^{(1)}_{nmp}(\mathbf{r},k_f)\otimes\left[(1-\delta_{s1})A_N^{fs}\mathbf{N}^{(0)}_{nmp}(\mathbf{r^\prime},k_s)\right.\nonumber\\
 &&\quad\quad\quad\quad \left.+(1-\delta_{sN})B^{fs}_N\mathbf{N}^{(1)}_{nmp}(\mathbf{r^\prime},k_s)\right]\nonumber\\
 &&+(1-\delta_{f1})\mathbf{M}^{(0)}_{nmp}(\mathbf{r},k_f)\otimes\left[(1-\delta_{s1})C_M^{fs}\mathbf{M}^{(0)}_{nmp}(\mathbf{r^\prime},k_s)\right.\nonumber\\
 &&\quad\quad\quad\quad \left.+(1-\delta_{sN})D^{fs}_M\mathbf{M}^{(1)}_{nmp}(\mathbf{r^\prime},k_s)\right]\nonumber\\
 &&+(1-\delta_{f1})\mathbf{N}^{(0)}_{nmp}(\mathbf{r},k_f)\otimes\left[(1-\delta_{s1})C_N^{fs}\mathbf{N}^{(0)}_{nmp}(\mathbf{r^\prime},k_s)\right.\nonumber\\
 &&\quad\quad\quad\quad \left.\left.+(1-\delta_{sN})D^{fs}_N\mathbf{N}^{(1)}_{nmp}(\mathbf{r^\prime},k_s)\right]\right\},\nonumber\label{eq:Green_scattered}\\
\end{eqnarray}
and the field and source wave numbers are
\begin{equation}
 k_{f,s}=\frac{\omega}{c}\sqrt{\mu_{f,s}\epsilon_{f,s}},
\end{equation}
$\mu_{f,s}$ and $\epsilon_{f,s}$ being the relative permeability and permittivity, in layer $s$ and $f$, respectively. The coefficients $A^{fs}_{M,N}$, $B^{fs}_{M,N}$, $C^{fs}_{M,N}$, and $D^{fs}_{M,N}$ are found by imposing the boundary conditions
\begin{eqnarray}
\begin{aligned}
 \lim_{\delta\rightarrow 0}\hat{\mathbf{r}}\times\bar{\bar{G}}_{r=R_j-\delta}&=\lim_{\delta\rightarrow 0}\hat{\mathbf{r}}\times\bar{\bar{G}}_{r=R_j+\delta}\\
 \frac{\lim_{\delta\rightarrow 0}}{\mu_j}\hat{\mathbf{r}}\times\boldsymbol{\nabla}\times\bar{\bar{G}}_{r=R_j-\delta}&=\frac{\lim_{\delta\rightarrow 0}}{\mu_{j+1}}\hat{\mathbf{r}}\times\boldsymbol{\nabla}\times\bar{\bar{G}}_{r=R_j+\delta},
 \end{aligned}
\end{eqnarray}
meaning that the tangential component of the electric and magnetic field is continuous as we approach the boundary between two layers with the field point $\mathbf{r}$ from two sides of the interface between layers $j$ and $j+1$. Solving for each boundary, one obtains the coefficients in $\bar{\bar{G}}_S$ and thus the total Green tensor.

\section{Dirac delta operator expanded on spherical vector harmonics}\label{sec:app_Dirac}

In order to perform the mode-selective quantization in the spherically symmetric system, we must also expand the singular term in (\ref{eq:app_G0}) in the spherical vector harmonic basis. To do so, we expand the total unit operator, choosing spherical Bessel functions for the radial parts, i.e., $z_n=j_n$. Because of the completeness of the basis, we can express the delta operator as
\begin{eqnarray}
 &&\bar{\bar{\delta}}(\mathbf{r}-\mathbf{r^\prime})=\!\!\sum_{nmp}\!\int_0^\infty\!\!\!\!\!\!\mathrm{d}q\!\left[A_{nmp}(q)\mathbf{M}^{(0)}_{nmp}(\mathbf{r},q)\otimes\mathbf{M}^{(0)}_{nmp}(\mathbf{r^\prime},q)\right.\nonumber \\
 &&\quad\quad\quad\quad\quad+B_{nmp}(q)\mathbf{N}^{(0)}_{nmp}(\mathbf{r},q)\otimes\mathbf{N}^{(0)}_{nmp}(\mathbf{r^\prime},q)\nonumber \\
 &&\quad\quad\quad\quad\quad+\left.C_{nmp}(q)\mathbf{L}^{(0)}_{nmp}(\mathbf{r},q)\otimes\mathbf{L}^{(0)}_{nmp}(\mathbf{r^\prime},q)\right],\nonumber \\
\end{eqnarray}
which, taking the notation of (\ref{eq:app_scalar_prod}), is the position representation of the unit operator expansion
\begin{eqnarray}
\begin{aligned}
 \hat{\mathbbm{1}}=&\sum_{nmp}\int_0^\infty\!\!\!\!\!\mathrm{d}q\big[A_{nmp}(q)|\mathbf{M}_{nmp}(q)\rangle\langle \mathbf{M}_{nmp}(q)|\\
 &+B_{nmp}(q)|\mathbf{N}_{nmp}(q)\rangle\langle \mathbf{N}_{nmp}(q)|\\
 &+C_{nmp}(q)|\mathbf{L}_{nmp}(q)\rangle\langle \mathbf{L}_{nmp}(q)|\big].\label{eq:app_1_expansion}
\end{aligned}
 \end{eqnarray}
In order to find $A_{nmp}(q)$, we multiply both sides with $|\mathbf{M}\rangle$ eigenfunctions:
\begin{eqnarray}
 \begin{aligned}
\langle\mathbf{M}_{n^\prime m^\prime p^\prime}(q^\prime)|\hat{\mathbbm{1}}|\mathbf{M}_{n^{\prime\prime} m^{\prime\prime} p^{\prime\prime}}(q^{\prime\prime})\rangle=\sum_{nmp}\int_0^\infty\!\!\!\!\!\mathrm{d}qA_{nmp}(q)\\
\times\langle\mathbf{M}_{n^\prime m^\prime p^\prime}(q^\prime)|\mathbf{M}_{nmp}(q)\rangle\langle \mathbf{M}_{nmp}(q)|\mathbf{M}_{n^{\prime\prime} m^{\prime\prime} p^{\prime\prime}}(q^{\prime\prime})\rangle.
 \end{aligned}
\end{eqnarray}
Using the orthogonality relations (\ref{eq:app_ortho2}) and performing the sums and integration we find
\begin{equation}
 A_{nm^e_o}(q)=\frac{q^2(2n+1)(n-m)!}{\pi^2 n(n+1)(n+m)!(1\pm \delta_{m0})},\label{eq:app_dirac_A}
\end{equation}
where the upper sign refers to $p=e$ and the lower one to $p=o$. It is easy to verify that
\begin{equation}
 \mathbf{M}_{n0o}(\mathbf{r},q)=\mathbf{N}_{n0o}(\mathbf{r},q)=\mathbf{L}_{n0o}(\mathbf{r},q)=\mathbf{0},
\end{equation}
that is, for $m=0$, $p=o$ we do not need a coefficient in the expansion. Thus, the divergence in the expression (\ref{eq:app_dirac_A}) does not pose a problem.

Repeating the procedure for the rest of the coefficients, we find
\begin{eqnarray}
&&\hat{\mathbbm{1}}=\sum_{n=0}^\infty\sum_{m=0}^n\int_0^\infty\!\!\!\!\!\mathrm{d}q\frac{q^2(2n+1)(n-m)!}{\pi^2n(n+1)(n+m)!(1+\delta_{m0})}\nonumber\\
&&\times\sum_{p=e,o}\big[|\mathbf{M}_{nmp}(q)\rangle\langle \mathbf{M}_{nmp}(q)|+|\mathbf{N}_{nmp}(q)\rangle\langle \mathbf{N}_{nmp}(q)|\nonumber\\
&&\quad\quad\quad+n(n+1)|\mathbf{L}_{nmp}(q)\rangle\langle \mathbf{L}_{nmp}(q)|\big],\nonumber\\
\end{eqnarray}
or, in position representation:
\begin{eqnarray}
&&\bar{\bar{\delta}}(\mathbf{r}-\mathbf{r^\prime})=\sum_{nmp}\int_0^\infty\!\!\!\!\!\mathrm{d}q\frac{q^2(2n+1)(n-m)!}{\pi^2n(n+1)(n+m)!(1+\delta_{m0})}\nonumber\\
&&\times\big[\mathbf{M}_{nmp}(\mathbf{r},q)\otimes \mathbf{M}_{nmp}(\mathbf{r^\prime},q)+\mathbf{N}_{nmp}(\mathbf{r},q)\otimes \mathbf{N}_{nmp}(\mathbf{r^\prime},q)|\nonumber\\ 
&&\quad\quad\quad+n(n+1)\mathbf{L}_{nmp}(\mathbf{r},q)\otimes \mathbf{L}_{nmp}(\mathbf{r^\prime}q)\big].\nonumber\\ \label{eq:app_Dirac_expansion}
\end{eqnarray}

To expand the singular term in $\bar{\bar{G}}_0$  (\ref{eq:app_G0}) one simply takes the $\hat{\mathbf{r}}\otimes\hat{\mathbf{r}}$ component of (\ref{eq:app_Dirac_expansion}).

\end{appendix}

\end{document}